\documentclass[aps,prd,preprint,showpacs,floatfix,preprintnumbers,nofootinbib,superscriptaddress,showkeys]{revtex4}
\usepackage[utf8]{inputenc}
\usepackage{float}
\usepackage{color}
\usepackage{fancybox}
\usepackage{hhline}
\usepackage{dcolumn}
\usepackage{textcomp}
\usepackage{epsfig,graphics,graphicx}
\usepackage{amsfonts,amssymb,amsmath}
\usepackage{pifont}
\usepackage{bm}
\usepackage{longtable} 
\usepackage{appendix}
\usepackage{lscape}
\usepackage[mathscr]{euscript}
\usepackage{mathrsfs}
\usepackage{multirow}
\usepackage{rotating}
\usepackage{color} 
\usepackage{placeins}
\usepackage[dvipsnames]{xcolor}

\newcommand{\beq}{\begin{equation}}
\newcommand{\eeq}{\end{equation}}
\newcommand{\bea}{\begin{eqnarray}}
\newcommand{\eea}{\end{eqnarray}}

\newcommand{\eps}{\epsilon}
\newcommand{\ord}[1]{{\cal{O}}\left( #1 \right)}

\DeclareFontFamily{OT1}{pzc}{}
\DeclareFontShape{OT1}{pzc}{m}{it}%
              {<-> s * [0.900] pzcmi7t}{}
\DeclareMathAlphabet{\mathpzc}{OT1}{pzc}%
                                 {m}{it}
\DeclareMathAlphabet{\mathcalligra}{T1}{calligra}{m}{n}
\usepackage{hyperref}
\hypersetup{pdfauthor={me}, colorlinks=true, citecolor=blue, urlcolor=blue, linkcolor=black}
\begin{document}
\preprint{\vbox{\hbox{ JLAB-THY-14-1961} }}
\title{\phantom{x}
\vspace{-0.5cm}     }
\title{Baryon spin-flavor structure  from an analysis of lattice QCD results of  the baryon spectrum}
\author{I. P. Fernando}\email{ishara@jlab.org }
%
\author{J.~L.~Goity}\email{goity@jlab.org}
\affiliation{Department of Physics, Hampton University, Hampton, VA 23668, USA. }
\affiliation{Thomas Jefferson National Accelerator Facility, Newport News, VA 23606, USA.}
\date{\today}
 
\begin{abstract} 
{ The excited   baryon masses are analyzed in the framework of the $1/N_c$ expansion  using the  available physical masses and also the masses obtained in lattice QCD   for different  quark masses. 
The baryon states are organized into irreducible representations of   $SU(6)\times O(3)$, where the $[{\bf{56}},\ell^P=0^+]$ ground state and excited baryons,  and    the $[{\bf{56}},2^+]$ and $[{\bf{70}},1^-]$ excited states are analyzed.
The analyses are carried out to  $\ord{1/N_c}$  and first order in the quark masses. The issue of state identifications is discussed.
Numerous parameter independent mass relations result at those orders, among them the well known Gell-Mann-Okubo and Equal Spacing   relations, as well as additional relations involving baryons with different spins. It is observed that such relations are satisfied at the expected level of precision.  From the quark mass dependence of the  coefficients in the baryon mass formulas  an increasingly simpler  picture of the spin-flavor composition of the baryons is observed with increaing quark masses, as measured by the number of significant mass operators.  }   
 \end{abstract}
 
\pacs{11.15-Pg, 11.30-Rd, 12.39-Fe, 14.20-Dh}
\keywords{Baryons, Lattice QCD,  Large N}

\maketitle

  
\section{Introduction}\label{sec:Intro} 
One of the most important present objectives in lattice QCD (LQCD) is the calculation of the light baryon spectrum, where in recent years substantive progress has been made.  The implementation of  optimized  baryon source operators \cite{WalkerLoud:2008bp,Bulava:2009jb,Edwards:2011jj,Edwards:2012fx} has enabled improved signals for excited baryons, leading to remarkable progress in identifying states by their quantum numbers and in the determination of their masses. In calculations performed with quark masses   corresponding to  $390 ~{\rm MeV}\leq M_\pi\leq 702 ~{\rm MeV}$,   the spectrum of non-strange baryons   \cite{Edwards:2011jj} and also of strange baryons \cite{Edwards:2012fx} were obtained. These calculations were performed on anisothropic lattices $16^3\times 128$ with a   gluon Symanzik-improved action with
tree-level tadpole-improved coefficients and an  anisotropic clover fermion action as explained in Ref. \cite{Lin:2008pr}. Although the effects of finite widths of the baryons are not yet implemented in these calculations, the results are very significant. The extraction of the baryonic resonance parameters (mass and width) by means of finite volume effects on the two body spectrum (e.g., $\pi N$) as it has been carried out  for the $\rho$ meson \cite{Dudek:2012xn}, in baryons is still to be fully implemented in a LQCD calculation. A nice  example of the latter  was shown in a  continuum Chiral Perturbation Theory study of those effects  for extracting the $\Delta$ resonance \cite{Bernard:2007cm}. The results used in this work pertain to the use of quasi-local baryon source/sink operators, which are not entirely sufficient for extracting the resonance parameters, and therefore the quoted masses will probably be (slightly) shifted in the more complete framework employing the finite volume effects. 
In fact, for  the LQCD states to be analyzed here,  the available phase space for the two body decay of the excited baryons is increasingly suppressed with increasing $M_\pi$,  which for the considered range of quark masses result  in  state widths   which are  significantly smaller than in the physical case.  An  estimate using the available phase space and the phenomenological widths gives widths  $\sim 50$  MeV or smaller for the S-wave decays and even smaller for P- and D- waves.
Thus, the present results of the LQCD baryon masses are expected to be very close to those one would obtain with the more complete method. 
 
 Although other recent  works on baryon LQCD spectroscopy have been carried out in Refs. \cite{WalkerLoud:2008bp,Bulava:2009jb,Engel:2013ig,Alexandrou:2013fsu,Alexandrou:2013fsu,Mahbub:2013ala,Mahbub:2012ri}, the present work will use the results obtained by the Jefferson Lab Lattice QCD Collaboration in  Refs.  \cite{Edwards:2011jj,Edwards:2012fx}. The study can similarly be applied to other results, in particular those of the BGR Collaboration \cite{Engel:2013ig} where the masses of the states analyzed here have been calculated.

A key observation from the analysis carried out in  \cite{Edwards:2011jj,Edwards:2012fx} is that  source/sink operators which, in the continuum limit,  are in irreducible representations of the spin-flavor and quark orbital angular momentum groups  $SU(2N_f)\times O(3)$ are very close to be at the optimum for the selective overlap with the baryon states.   This  is a strong indication that the baryon mass eigenstates  themselves  must be approximately organized    into irreducible  multiplets  of that group, a fact that is  well known to hold phenomenologically. This has been tested explicitly in  the LQCD calculations by  measuring    the coupling strengths  of sources in different representations to each of the baryon levels studied.  The state admixture of different $SU(2N_f)\times O(3)$ irreducible representations, known as  {\it configuration mixing}, cannot however be directly inferred from those strengths, as they depend on details of the operators.  Since in the exact symmetry limit the couplings would be ''diagonal'', it is expected that the existence of small off diagonal couplings necessarily translates into small state configuration mixings. In the present work,  configuration mixings  will be  altogether neglected, and thus all claims are restricted to the approximate validity of that assumption. The states studied in this work are the ones corresponding to the   $SU(6)\times O(3)^P$ $[{\mathbf{56}},\,0^+]$ or Roper multiplet, the $[{\mathbf{56}},  2^+]$ and the $[{\mathbf{70}}, 1^-]$. These are of particular interest because they have been previously analyzed phenomenologically in the framework of the $1/N_c$ expansion employed here \cite{Matagne:2014lla}, where the assumption of no configuration mixing works very well up to the degree of accuracy of the input masses and other observables permit.  

The existence of a spin-flavor symmetry in baryons can be rigorously justified in the large $N_c$ limit of QCD. The  symmetry is the result of a consistency requirement imposed by unitarity on pion-baryon scattering in that limit \cite{Gervais:1983wq,Gervais:1984rc,Dashen:1993as}, and  spin-flavor symmetry is thus broken by corrections which can be organized in powers of $1/N_c$.  Under the assumption that the real world  with $N_c=3$ baryons can be analyzed using a $1/N_c$ expansion,  starting at the lowest order with an exact spin-flavor symmetry, many analyses of baryon masses and other properties have been carried out. In particular, excited baryon masses have been analyzed in numerous works for the cases considered in this work \cite{ Goity:1996hk,Pirjol:1997sr,Carlson:1998gw,Carlson:1998vx,Schat:2001xr,Goity:2002pu,Goity:2003ab,Pirjol:2003ye,deUrreta:2013koa} as well as for other multiplets \cite{Matagne:2004pm,Matagne:2005gd,Matagne:2006zf,Matagne:2012tm}. 
Although spin-flavor symmetry is justified in the large $N_c$ limit, the larger $SU(2N_f)\times O(3)$ is not. The latter can be broken due to spin-orbit effects at $\ord{N_c^0}$, as it is the case in the $[{\mathbf{70}}, 1^-]$ baryons, and is in principle not such a good symmetry. However, phenomenologically it has been known since old times that spin-orbit effects in baryons are small, and actually smaller than the hyperfine (HF) effects which are sub-leading in $1/N_c$.  In addition, configuration mixings which are not suppressed in the large $N_c$ limit turn out to be driven by operators of the spin-orbit type \cite{Goity:2004pw,Goity:2005gs}, and seem to be small as well. As mentioned earlier, these observations also apply to the LQCD baryons.

Particular predictions  result when  configuration mixings are disregarded. They have the form  of  parameter independent mass relations which hold up to higher order corrections in the $1/N_c$ or $SU(3)$ breaking expansions.
 Among those relations are the well known  Gell-Mann-Okubo (GMO) and equal spacing (EQS) relations, which are valid in general, and  additional ones  involving different spin-flavor states such as   relations in the 56-plets  that follow from the G\"ursey-Radicati mass formula, and other relations in the 70-plet \cite{Goity:2002pu}. As it will be shown in the present analysis,  LQCD baryon masses fulfill to the expected accuracy those relations. 

The objective of this work is to analyze the LQCD results for baryon masses using the $1/N_c$ expansion to $\ord{1/N_c}$ and to first order in $SU(3)$ symmetry breaking. Although the LQCD  results, as mentioned above,    are at larger than physical quark masses and do not have a complete implementation of the effects due to the finite  baryon decay widths, they provide complete sets of states, i.e., states that complete the experimentally partially filled multiplets, which is  a very useful addition for  more accurate analyses as the ones carried out here. In addition, since the $1/N_c$ expansion of QCD applies even in cases where such approximations are made (e.g., quenched QCD, larger quark masses, etc.), the present study   also serves   as a test of the $1/N_c$ expansion itself.

In the phenomenological analyses, the excited baryon   masses  used as inputs are those provided by the Particle Data Group (PDG) \cite{Beringer:1900zz}.  For two flavors and the multiplets considered here all states are established, but for three flavors   there is a significant number of missing strange baryon states. For example,  in the   $[{\mathbf{70}}, 1^-]$ multiplet  there are 30 theoretical masses and only 17   are currently experimentally known. Although those 17 masses are sufficient for the purpose of the $1/N_c$ analysis, they are not sufficient   for a thorough test of the mass relations.   On the other hand, the LQCD results  provide complete multiplets, enabling a complete test of mass relations.  In the particular case of the   $[{\mathbf{70}}, 1^-]$, the issue of state mixing can be sorted out in the phenomenological case thanks to the simultaneous analysis of partial decay widths and photo-couplings, as shown most recently in Ref. \cite{deUrreta:2013koa} for the non-strange baryons. These inputs are however not possible for the LQCD baryons, and therefore the state mixing relies very strongly on the criterion for identifying states. In this regard,  level crossing effects are possible as the quark masses are varied in the LQCD calculations \cite{Engel:2013ig,Mahbub:2013ala}. This is a present topic of interest in LQCD, which is still in its early stages in the study of the baryon spectrum.


This work is organized as follows: In Section \ref{sectionII} a brief description of the $1/N_c$ expansion framework is given;  Section III contains the  results and their analysis;  Section IV gives the summary and conclusions.   Appendix A displays the bases of operators and the respective matrix elements needed in this work, and Appendix B  gives the baryon masses, both from the  PDG  \cite{Beringer:1900zz} and  LQCD   \cite{Edwards:2011jj,Edwards:2012fx}, which are the inputs for the fits.

 \section{The  $1/N_c$ expansion and spin flavor symmetry in baryons}\label{sectionII}

Consistency of baryons in the ordinary large  $N_c$ limit as defined by 'tHooft \cite{'tHooft:1973jz} requires that baryons form multiplets of a contracted spin-flavor group $SU(2N_f)$ \cite{Gervais:1983wq,Gervais:1984rc,Dashen:1993as}. The generators of that symmetry are denoted by $S^i$ (spin), $T^a$ (flavor) and $X^{ia}=G^{ia}/N_c$ (spin-flavor). In the case of excited baryons the observation that quark  spin and orbital angular momentum are weakly coupled in baryons has lead to a phenomenologically successful scheme of organizing the states in multiplets of $SU(2N_f))\times O(3)$. Without loss of generality it is possible to work with the ordinary rather than the contracted spin-flavor group for the purposes of building the operator bases \cite{Dashen:1994qi}. 
 Any  static baryonic observable   can be expressed by an effective operator which is decomposed in a basis of operators ordered in powers of $1/N_c$ and which can be expressed as appropriate tensor products of the symmetry generators. In the present case  of baryon masses,  the bases of operators are well known.  The details for  obtaining those  bases can be found in Refs. \cite{Dashen:1993jt,Goity:1996hk,Goity200383,Carlson:1998gw,Carlson:1998vx,Schat:2001xr,Goity:2002pu}.


The excited states considered here will be either in the totally symmetric (Sym) or in the mixed symmetric   (MSym) irreducible representations of $SU(6)$.   Following the large $N_c$  Hartree picture of a baryon, without a loss of generality and for the purpose of dealing with the group theory of the spin-flavor and orbital degrees of freedom,  one can  describe a low excitation  baryon as a spin-flavor symmetric core with $N_c-1$  quarks and one excited  quark.  In this way it becomes straightforward to obtain the matrix elements of bases operators.  Appendix I gives the    mass operator bases to the needed order  and the corresponding matrix elements. 


%


The mass operator bases  are organized in powers  of $1/N_c$ and involve $SU(3)$ singlet and octet  operators, the latter for symmetry breaking by the parameter $\epsilon\equiv m_s-\hat{m}$, where $\hat{m}=(m_u+m_d)/2$. One may consider the expansion to $\ord{\epsilon^0/N_c }$ and $\ord{\epsilon}$. It turns out that contributions $\ord{\epsilon/N_c}$ are almost insignificant in most cases as  shown  later.

The multiplets to be analyzed have the following state contents: i) $[{\bf{56}},0^+]$: one $SU(3)$ $\bf 8$ with $S=1/2$ and one $\bf{10}$ with $S=3/2$; ii)  $[{\bf{56}},2^+]$: one $\bf 8$ for each $S=3/2$ and 5/2, and one $\bf{10}$ for each $S=1/2$ through 7/2; iii) $[{\bf{70}},1^-]$: one $\bf 1$ $\Lambda$ baryon for each $S=1/2$ and $3/2$, two $\bf 8$s for each $S=1/2$ and 3/2, one $\bf 8$ for $S=5/2$ and one $\bf{10}$ for each $S=1/2$ and $3/2$. 

For each case, the mass operators to the order needed here are as follows:

 $[{\bf{56}},0^+]$:   in this case the mass operator is the famous G\"ursey-Radicati (GR) mass formula, which, explicitly displaying the $1/N_c$ power counting,   reads as follows:
\bea
M_{[{\bf{56}},0^+]}&=&c_1N_c+\frac{c_2}{N_c}\;S(S+1) + 
b_1 N_s \nonumber \\ &  +& \dfrac{  b_2}{2\sqrt{12}\, N_c}\left(3I(I+1)-S(S+1)-\frac{3}{4}N_s(N_s+2)\right)+{\cal{O}}( 1/N_c^{2})\, , 
\eea
where $S$ is the baryon spin operator, $I$ the isospin, and $N_s$ the number of strange quarks, and the $c_i$ and $b_i$ are coefficients determined by the QCD dynamics,   which are obtained by fitting to the masses. The mass operators as defined such that all coefficients are $\ord{N_c^0}$.
The $SU(3)$ breaking parameter $\eps$ is here included in the coefficients $b_1$ and $b_2$.   For all mass formulas, the quark mass dependencies  are implicitly absorbed into the   coefficients.

 $[{\bf{56}},2^+]$: in this case the basis has  three $SU(3)$ symmetric  and three breaking operators:
 \beq
M_{[{\bf{56}},2^+]}=\sum_{i=1}^{3}c_{i}\,O_i+ \sum_{i=1}^3b_{i}\,{\bar B}_i\, ,
\eeq
The basis of operators along with the matrix elements are given in Appendix A1,  Table \ref{562plusbasis}.

$[{\bf{70}},1^-]$: In the case of non-strange baryons, where the states belong to a $\bf{20}$ plet of $SU(4)$ the mass formula reads \cite{Carlson:1998vx}:
\beq
M_{[{\bf{20}},1^-]}=\sum_{i=1}^8c_{i}\,O_i\, ,
\eeq
where the eight basis operators up to and including $\ord{1/N_c}$ are given in Table \ref{201minusbasis} of Appendix A2.
For three flavors the mass formula reads \cite{Schat:2001xr,Goity:2002pu}:
\beq
M_{[{\bf{70}},1^-]}=\sum_{i=1}^{11}c_{i}\,O_i+ \sum_{i=1}^4  b_{i}\,{\bar B}_i\, ,
\eeq
where the    basis operators up to and including $\ord{1/N_c}$ or $\ord{\eps}$  are given in Table \ref{701minusbasis} and Table \ref{701minus8basis} of Appendix A2. In order that the $SU(3)$ breaking operators do not contribute to the non-strange baryon masses, they have ben redefined according to:    $\bar{B_1}=t_8-\frac{1}{2\sqrt{3}N_c}O_1$,  $\bar{B_2}=T^c_8-\frac{N_c-1}{2\sqrt{3}N_c}O_1$, $\bar{B_3}=\frac{10}{N_c}\,d_{8ab}\,g_{ia}G_{ib}^c+\frac{5(N_c^2-9)}{8\sqrt{3}N_c^2(N_c-1)}O_1+\frac{5}{2\sqrt{3}(N_c-1)}O_6+\frac{5}{6\sqrt{3}}O_7$, $\bar{B_4}=3\,\ell_i\,g_{i8}-\frac{\sqrt{3}}{2}O_2$.

Since in general the number of states is larger than the number of coefficients of the fit, and the masses are linear in the coefficients,  there must be linear mass relations which are independent of the coefficients. Such mass relations have been derived in previous works, and will be tested here  with the LQCD results. Many of the mass relations involve $SU(3)$ breaking mass differences, and are thus identically satisfied in the limit of $SU(3)$ symmetry.  There are however some mass relations which test  exclusively the breaking of the spin-symmetry at $\ord{1/N_c}$; these occur in the ${[{\bf{56}},2^+]}$ multiplet. The mass relations will be presented in the analysis of each case below, and 
they  are depicted in  Tables \ref{MassRel1},  \ref{MassRel2},  \ref{MassRel3},  \ref{MassRel4},  \ref{MassRel5}.

\section{Fits to the LQCD results}

In this section the fits to the LQCD masses  are performed.  The LQCD results used here are as follows: for two flavors the results are those of Ref. \cite{Edwards:2011jj}, of which only the results for the negative parity baryon masses will be analyzed, and for three flavors the results of Ref. \cite{Edwards:2012fx} are used. For two flavors the quark masses used correspond to $M_\pi=396$ and 524 MeV, and for three flavors $m_s$ has been kept fixed, and $M_\pi=391$, 524 and 702 MeV, where the last one corresponds to exact $SU(3)$ symmetry.

For each of the multiplets it is necessary to identify the states with the LQCD mass levels. This procedure is not unique and thus it requires some analysis, as shown below.  In the following the notation used to designate the states will be as follows: $B_S$ or $B'_S$   for states with baryon spin $S$ which belong predominantly in octets, and $B^{''}_S$ for baryons which belong predominantly in singlet or decuplet. For the case of the $\Delta$ and $\Omega$ baryons which can only belong in a decuplet when isospin is exact, no primes are used, and the same for the $[\mathbf{56},0^+]$ baryons where $S=1/2 ~(3/2)$ states necessarily belong to  ${\bf 8}~({\bf 10})$.




\subsection{$[\mathbf{56},0^+]$    Baryons}






Here, the   ground state and excited Roper  baryon masses are fitted  using the GR mass formula Eq. (1). In all ${\bf 56}$-plet masses the flavor singlet breaking of $SU(6)\times O(3)$ is $\ord{1/N_c}$, i.e., suppressed by a factor $1/N_c^2$ with respect to the leading symmetric mass. Thus, under the assumption of no configuration mixing,  $SU(6)\times O(3)$  must be particularly good.  The possible significance of the $SU(3)$ breaking effects on the HF terms, controlled by the coefficient $b_2$,  is  considered.  Table \ref{FitGS} gives the results of the fits for the ground state baryons, and Table \ref{FitRoper} for the Roper baryons.  

The analysis of LQCD ground state baryon masses including higher order terms in the $SU(3)$ breaking has been carried out in Ref. \cite{Jenkins:2009wv}, for   LQCD calculations other than the present ones.
It is noted that the HF mass splittings have the behavior observed also in other LQCD calculations, where it increases with $M_\pi$ up to $M_\pi\sim 400$ MeV, to decrease for higher $M_\pi$ (for a current summary see Ref. \cite{Cordon:2014sda}). On the other hand for the excited  baryons the HF splittings are almost always monotonously decreasing with increasing $M_\pi$, both in the  ${\bf 56}$- and  ${\bf 70}$-plets.

In  the Roper baryons,  the identification of the $\bf {8}_{1/2}$ is obvious, being the lightest positive parity excited states above the ground state, but for the $\bf{10}_{3/2}$ one needs to distinguish between two possible excited multiplets, one which will be a Roper and one which be in  the $\mathbf{[{\bf{56}},2^+]}$.  One of the choices, namely that in which the Roper $\bf{10}_{3/2}$ is the lightest one, does not seem to be consistent with the magnitude of the HF splittings observed throughout the spectrum. One is therefore lead to conclude that the   $\bf{10}_{3/2}$ belonging to the  $\mathbf{[{\bf{56}},2^+]}$ are the lowest lying excitations, followed by the Roper ones.

\begin{table*}[htpb!!!]
\begin{tabular}{lcccc }
\hline 
 \hline Coefficients & &  &  $M_\pi [MeV]$& \\
 $[{\rm MeV}]$&  ~~~~PDG &~~~~$391$ & ~~~~$524$ & ~~~~$702$    \\ 
\hline $c_1 $ & ~~~~293$\pm$6 & ~~~~377$\pm$3 & ~~~~420$\pm$2 & ~~~~474 $\pm$1    \\ 
 $c_2 $ & ~~~~247$\pm$12 & ~~~~296$\pm$5 & ~~~~251$\pm$3 & ~~~~200$\pm$2   \\ 
 $b_1 $ & ~~~~189$\pm$12 & ~~~~75$\pm$6 & ~~~~45$\pm$3 & ~~~~0      \\ 
 $b_2 $ & ~~~~94$\pm$26 & ~~~~43$\pm$11 & ~~~~14$\pm$7 & ~~~~0  \\ 
\hline $\chi^2_{\rm dof}$ & ~~~~0.19 & ~~~~0.15 & ~~~~1.43 & ~~~~0    \\ 
\hline \hline
\end{tabular}  
\caption{Coefficients of the GR mass formula for the ground state baryons.  The case $M_\pi=702$ MeV corresponds to exact $SU(3)$ symmetry.   $\chi^2_{\rm dof}$ is the $\chi^2$ per degree of freedom.}
\label{FitGS}
\end{table*}


\begin{table*}[t]
\begin{tabular}{lccccc}\hline
\hline Coefficients & & $M_\pi [MeV]$ & &  & \\
 $[{\rm MeV}]$ & ~~~~PDG & ~~~~391 & ~~~~524 & ~~~~702 \\ 
\hline $c_1 $ & ~~~~469$\pm$8 & ~~~~714$\pm$6& ~~~~760$\pm$5& ~~~~770$\pm$3 \\ 
 $c_2 $ & ~~~~175$\pm$44 & ~~~~165$\pm$12 & ~~~~124$\pm$9 & ~~~~115$\pm$20 \\ 
 $b_1 $ & ~~~~204$\pm$18 & ~~~~48$\pm$12 & ~~~~15$\pm$12 &~~~~0 \\ 
\hline $\chi^2_{\rm dof}$ & ~~~~0.16 & ~~~~0.53 & ~~~~0.76 & ~~~~0 \\ 
\hline \hline
\end{tabular}  
\caption{ Fit to the  $[{\mathbf{56}},\,0^+]$ excited Roper baryons. It is found that  the $SU(3)$ breaking effects on the HF interactions can be neglected, thus $b_2=0$  throughout.}
\label{FitRoper}
\end{table*}



In Fig. 1 the dependencies on $M_\pi$ of the coefficients are displayed. The well known dramatic downturn with decreasing $M_\pi$ of the Roper baryon masses is clearly driven by the spin-flavor singlet component of the masses, given by the coefficient $c_1$. The HF effects  determined by $c_2$ have a smooth behavior in $M_\pi$ but significantly different strength in the GS than in the Roper states, being reduced in the latter.  Unlike the GS baryons, no significant  $SU(3)$ breaking in the HF interaction is observed in the Roper baryons, so the coefficient $b_2$ is consistent with zero for the LQCD masses.

\begin{figure}[h!]
\centering
\includegraphics[width=0.4\linewidth]{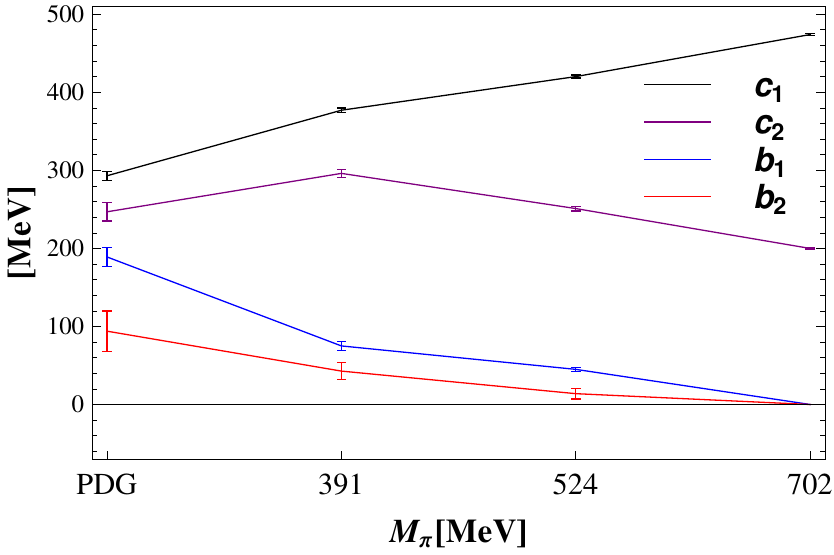}~~~~~~~\includegraphics[width=0.4\linewidth]{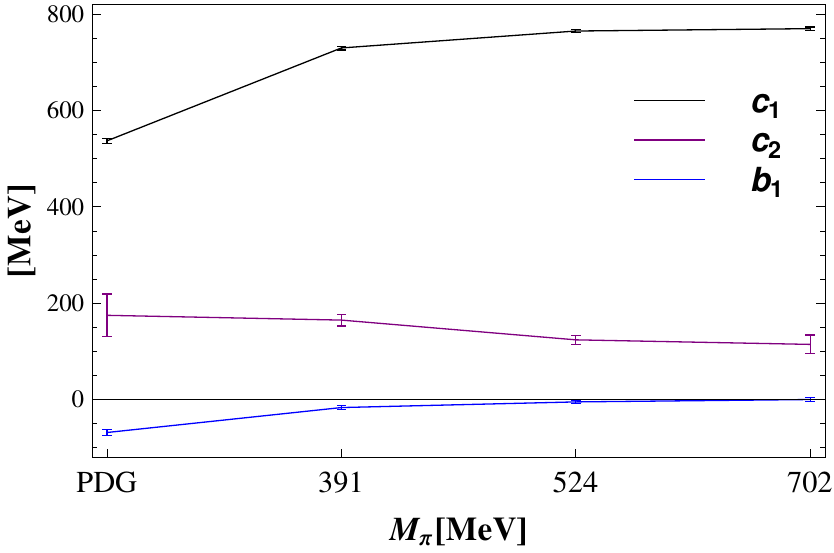}
\caption{
Evolution of the coefficients with $M_\pi$ for the ground state baryons (left panel) and the Roper baryons (right panel). 
} 
\label{fig:GroundStEvolution3parameterNc}
\end{figure}




The mass relations are given in Tables \ref{MassRel1} and \ref{MassRel2},  which show  that they are satisfied within errors for the LQCD results.  In the physical case,  the knowledge of the Roper states is  rather  incomplete. Based on the mass relations the  predictions shown in Table \ref{RoperPredictions} are made. As shown below, the listed PDG candidate states may also match predictions from the  $[\mathbf{56},2^+]$ multiplet, as discussed later.

\begin{table}[t]
\begin{tabular}{l c c c c}
\hline\hline Relation & &  & $M_\pi$[MeV] &\\
  &   & PDG~~ & 391~~ & 524~~ \\ 
\hline
$2(N+\Xi)-(3\Lambda+\Sigma)=0$ & & 30.2$\pm$0.4~~ & 38$\pm$75~~ & 32$\pm$32  \\
$\Sigma''-\Delta=\Xi''-\Sigma''=\Omega''-\Xi''$ & & 155$\pm$2~~ & 64$\pm$25~~  & 40$\pm$11  \\
& & 149.0$\pm$0.5~~ & 55$\pm$19~~ & 33$\pm$13  \\
& & 140.7$\pm$0.5~~ & 54$\pm$17~~ & 40$\pm$14  \\
$\frac{1}{3}(\Sigma+2\Sigma'')-\Lambda-(\frac{2}{3}(\Delta-N))=0$ & & 9$\pm$1~~ & 1$\pm$28~~ & 14$\pm$12 \\ 
$\Sigma''-\Sigma-(\Xi''-\Xi)=0$ & & 23.5$\pm$0.5~~ & 12$\pm$25~~ & 12$\pm$15 \\ 
$3\Lambda+\Sigma-2(N+\Xi)+(\Omega-\Xi''-\Sigma''+\Delta)=0$ & & 16$\pm$2~~ & 29$\pm$81~~ & 32$\pm$36 \\ 
$\Sigma''-\Delta+\Omega-\Xi''-2(\Xi''-\Sigma'')=0$ & & 2.5$\pm$2.4~~ &  8$\pm$51~~ & 14$\pm$37 \\ 
\hline\hline
\end{tabular} 
\caption{Mass relations for the ground state octet and decuplet. The relations are valid  up to corrections $\ord{\epsilon^2/N_c}$ in the case of the GMO and EQS relations, and up to  $\ord{\epsilon/N_c^2}$ for the rest. } 
\label{MassRel1}
\end{table}

\begin{table*}[t]
\begin{tabular}{l c c c}
\hline \hline Relation & &   ~~~~~~~~~~~~~~~~~~~~~~~~~~~$M_\pi$[MeV] & \\
 &    & 391~~ & 524  \\ 
\hline
$2(N+\Xi)-(3\Lambda+\Sigma)=0$ &&  179$\pm$180~~ & 106$\pm$155 \\
$\Sigma''-\Delta=\Xi''-\Sigma''=\Omega''-\Xi''$ &&  13$\pm$45~~  & -27$\pm$26 \\
&&  84$\pm$40~~ & 41$\pm$49  \\
&  & 48$\pm$42~~ & 41$\pm$57  \\
$\frac{1}{3}(\Sigma+2\Sigma'')-\Lambda-(\frac{2}{3}(\Delta-N))=0$ &&  51$\pm$65~~ & 29$\pm$41 \\ 
$\Sigma''-\Sigma=\Xi''-\Xi$ & &  58$\pm$63~~ & 77$\pm$80 \\ 
$3\Lambda+\Sigma-2(N+\Xi)+(\Omega''-\Xi''-\Sigma''+\Delta)=0$ &   & 144$\pm$189~~ & 174$\pm$170\\ 
$\Sigma''-\Delta+\Omega''-\Xi''-2(\Xi^*-\Sigma'')=0$ &&    107$\pm$110~~ & 67$\pm$147 \\ 
\hline \hline
\end{tabular} 
\caption{Mass relations for the   Roper multiplet. The relations hold at the same orders as in the case of the ground state baryons. } 
\label{MassRel2}
\end{table*}


\begin{table*}[h!!!!!!!!!!]
\begin{tabular}{ lccc }
\hline \hline Baryon~~~~~~ &  Predicted mass [MeV] & ~~~Fitted Mass [MeV] &   PDG candidate and mass [MeV]
  \\ 
\hline
$ \Sigma_{3/2}^{''}$& $1790\pm131$& 1800 & $\Sigma(1840)(3/2^+)^*~\text{with mass}\sim\,{1840}$ \\
$\Xi_{1/2}$& $1825\pm108 $& 1815 & $\cdots$\\
$\Xi_{3/2}^{''}$& $1955\pm171$& 1975 & $~~~~~  \Xi(1950)(?^?)^{***} ~\text{with mass}\sim 1950\pm15$\\
$\Omega_{3/2}$& $2120\pm 219$& 2150 & $\cdots$\\
\hline \hline
\end{tabular} 
\caption{Predictions of physically unknown states in the   Roper multiplet. These predictions agree with the ones in Ref. \cite{Carlson:2000zr}.  } 
\label{RoperPredictions}
\end{table*}


\newpage

~~~

\subsection{$[{\mathbf{56}}, 2^+]$ Baryons} 
 Here  the lowest excited baryons that can fit into the $[{\mathbf{56}}, 2^+]$ are considered.   The first step is the identification of the states in the LQCD results.  With the exception of the $\bf{10}_{3/2}$, all the states are in spin-flavor states which appear for the first time, and thus the lightest states with given spin and flavor are the ones of interest. In the case of the   $\bf{ 10}_{3/2}$, as discussed earlier, there are two possible excited levels to consider, one of which will belong to the excited  $[{\mathbf{56}}, 0^+]$, where the  arguments for the identification were already given.    For $\Sigma$ and $\Xi$ baryons, the LQCD analysis \cite{Edwards:2012fx} has assigned the dominant $SU(3)$ multiplet to which they belong, $\bf 8$ or $\bf 10$. Therefore, there is no ambiguity about the identification of states in the present multiplet.

 There is  mixing  between states in the octet and decuplet, namely the $\Sigma$ and the $\Xi$ pairs of states with $S=3/2$ and with $S=5/2$, namely 
 $(\Sigma^{(8)}_S$, $\Sigma^{(10)}_S)$   and $(\Xi^{(8)}_S$, $\Xi^{(10)}_S)$. These mixings obviously result from $SU(3)$ breaking, and the physical states are defined as follows:\\
\bea
\begin{pmatrix}
\Sigma_S \\ \Sigma'_S
\end{pmatrix}=\begin{pmatrix}
\cos \theta_{\Sigma_S} & \sin \theta_{\Sigma_S}  \\ - \sin \theta_{\Sigma_S}  & \cos \theta_{\Sigma_S} 
\end{pmatrix} \begin{pmatrix}
\Sigma^{(8)}_S \\ \Sigma^{(10)}_S
\end{pmatrix}~~,~~~
\begin{pmatrix}
\Xi_S \\ \Xi'_S
\end{pmatrix}=\begin{pmatrix}
\cos \theta_{\Xi_S}  & \sin \theta_{\Xi_S} \\ - \sin \theta_{\Xi_S} & \cos \theta_{\Xi_S}
\end{pmatrix} \begin{pmatrix}
\Xi^{(8)}_S \\ \Xi^{(10)}_S
\end{pmatrix} 
\eea
Two different fits are carried out, one includes all the $SU(3)$ breaking operators, and a  second one   only including  the  one-body  operator giving  the spin independent breaking effects. Since the symmetry breaking by the operator $B_1$ does not produce mixing between $\bf{8}$ and $\bf{10}$, the mixing angles are actually $\propto \eps/N_c$, and thus naturally very small. The results are shown in Table \ref{Fit562plus}. It is checked that the present fit fully agrees with a previous one for the physical case \cite{Goity:2003ab}.  One important observation is that based on the quality of the fits   the mixings cannot be definitely established for the LQCD results.
\begin{table*}[h!]
 \begin{tabular}{l c c c c c c c}
 \hline \hline Coefficients & & & & $M_\pi [MeV]$ & & & \\
 $[{\rm MeV}]$    & PDG & 391  & 524  & 702  & 391 & 524 & 702\\ 
\hline $c_1$ & 540$\pm$11 & 704$\pm$2 & 718$\pm$1 & 754$\pm$1 & 710$\pm$2~~~ & 724$\pm$1~~~ & 753$\pm$1 \\ 
 $c_2$ & 18$\pm$5 & 48$\pm$6 & 28$\pm$3 & -6$\pm$5 & 59$\pm$6~~~ & 21$\pm$3~~~ & 0 \\ 
 $c_3$ & 244$\pm$4 & 169$\pm$5 & 166$\pm$3 & 104$\pm$4 & 151$\pm$5~~~ & 148$\pm$3~~~ & 106$\pm$4 \\ 
 $b_1$ & 217$\pm$4 & 75$\pm$3 & 54$\pm$1 & 0& 56$\pm$3~~~ & 36$\pm$1~~~ & 0  \\ 
 $b_2$ & 95$\pm$14 & -23$\pm$11 & 13$\pm$5 & 0 & 0 & 0 & 0 \\ 
 $b_3$ & 268$\pm$16 & 59$\pm$9 & 55$\pm$4 & 0 & 0 & 0 & 0\\
\hline Mixing angles $[\text{Rad}]$& & & &  & & & \\
\hline $\theta_{\Sigma_{3/2}}$ & -0.16$\pm$0.02~~~~ & 0.06$\pm$0.03~~~~ & -0.03$\pm$0.01~~~~ & 0 & 0 & 0 & 0 \\ 
 $\theta_{\Xi_{3/2}}$ & -0.26$\pm$0.04~~~~ & 0.07$\pm$0.03~~~~ & -0.03$\pm$0.01~~~~ & 0 & 0 & 0 & 0 \\ 
 $\theta_{\Sigma_{5/2}}$ & -0.22$\pm$0.03~~~~ & 0.05$\pm$0.03~~~~ & -0.03$\pm$0.01~~~~ & 0 & 0 & 0 & 0 \\ 
 $\theta_{\Xi_{5/2}}$ & -0.20$\pm$0.02~~~~ & 0.08$\pm$0.04~~~~ & -0.03$\pm$0.01~~~~ & 0 & 0 & 0 & 0 \\ 
\hline $\chi^2_{\rm dof}$ & 0.84& 0.60 & 0.47 & 0.92  & 0.63 & 0.53 & 0.80\\ 
\hline \hline
\end{tabular} 
\caption{Two fits, with and without  the operators $B_2$ and $B_3$.  The second fit does not describe well the physical baryons. 
}
\label{Fit562plus}
\end{table*}
~~

\begin{figure}[h!!!!!]
\centering
\includegraphics[width=0.45\linewidth]{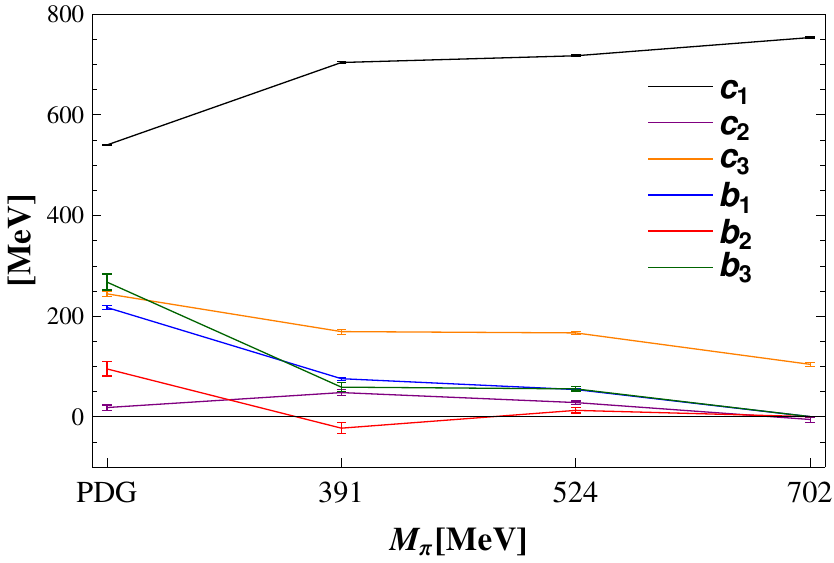}~~~~~\includegraphics[width=0.45\linewidth]{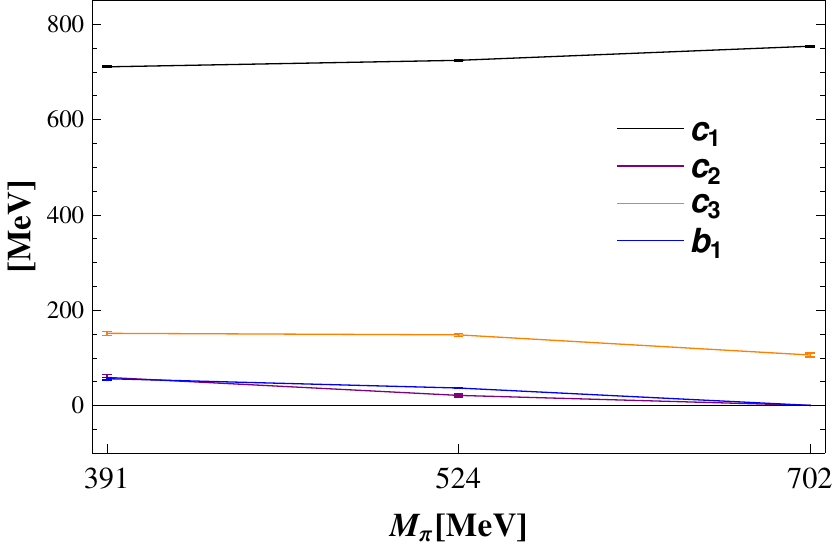}
\caption{Evolution of the operator coefficients with $M_\pi$ for the two fits of the  $[{\mathbf{56}}, 2^+]$ masses. }
\label{fig:[56,2+]withAllcoef}
\end{figure}

The evolution with $M_\pi$ of the coefficients is shown in Fig. 2. It is interesting to notice that the coefficient $c_1$ has  qualitatively similar but less dramatic behavior than in the case of the Roper baryons, which must  be an indication of a similar mechanism as the one which drives down the Roper masses with decreasing $M_\pi$. 
   The HF interaction given by  $c_3$ behaves smoothly with $M_\pi$,  decreasing slowly as   $M_{\pi}$ increases, and it has similar strength as in  the Roper baryons.   Although the operators $B_2$ and $B_3$ are significant  in the physical case, their contributions  are  negligible in the LCQD cases, as shown by the second fit in Table \ref{Fit562plus}. The latter observation tells that the mixing between octet and decuplet states, which are driven by those operators,  are very small as  confirmed by the small mixing angles in the first fit  in Table \ref{Fit562plus}.  

The mass relations for the  $[{\mathbf{56}}, 2^+]$   are depicted in Table \ref{MassRel3} \cite{Goity200383}. In addition to GMO and EQS relations,  there are several relations which relate $SU(3)$ mass splittings in multiplets with different baryon spin, as well as relations among the masses of baryons with the same strangeness but different baryon spin. Almost all relations are satisfied by the LQCD results, with the exception of the results at $M_\pi=702$ MeV, where the deviations are however within the expected magnitude of higher order corrections.

\begin{table*}[htpb!!!]
\renewcommand{\tabcolsep}{.25cm}
{\squeezetable
\begin{tabular}{l c c c}
\hline\hline
 Relation & & $M_\pi$[MeV]& \\
  & 391 & 524 & 702\\ 
\hline
$2(N_{3/2}+\Xi_{3/2})-(3\Lambda_{3/2}+\Sigma_{3/2})=0$ & 98$\pm$126 & 49$\pm$173 & 0 \\
$2(N_{5/2}+\Xi_{5/2})-(3\Lambda_{5/2}+\Sigma_{5/2})=0$ & 40$\pm$98 & 55$\pm$65 & 0 \\
$\Sigma_{1/2}''-\Delta_{1/2}=\Xi_{1/2}''-\Sigma_{1/2}''=\Omega_{1/2}-\Xi_{1/2}''$ & -13$\pm$110 & 36$\pm$33 & 0 \\
 & 23$\pm$44 & 43$\pm$22  & 0 \\
 & 85$\pm$54 & 35$\pm$19  & 0 \\
$\Sigma_{3/2}''-\Delta_{3/2}=\Xi_{3/2}''-\Sigma_{3/2}''=\Omega_{3/2}-\Xi_{1/2}''$ & 48$\pm$46 & 36$\pm$23 & 0 \\
 & 56$\pm$29 & 30$\pm$16  & 0 \\
 & 45$\pm$31 & 41$\pm$15  & 0 \\
$\Sigma_{5/2}''-\Delta_{5/2}=\Xi_{5/2}''-\Sigma_{5/2}''=\Omega_{5/2}-\Xi_{5/2}''$ & 35$\pm$40 & 34$\pm$26 & 0 \\
 & 62$\pm$31 & 26$\pm$23  & 0 \\
 & 57$\pm$34 & 52$\pm$18  & 0 \\
$\Sigma_{7/2}''-\Delta_{7/2}=\Xi_{7/2}''-\Sigma_{7/2}''=\Omega_{7/2}-\Xi_{7/2}''$ & 38$\pm$38 & 35$\pm$25 & 0 \\
 & 67$\pm$31 & 36$\pm$20  & 0 \\
 & 59$\pm$31 & 22$\pm$18  & 0 \\
$\Delta_{5/2}-\Delta_{3/2}-(N_{5/2}-N_{3/2})=0$ & 70$\pm$68 & 4$\pm$68 & 44$\pm$33 \\
$(\Delta_{7/2}-\Delta_{5/2})-\frac75(N_{5/2}-N_{3/2})=0$ & 68$\pm$78 & 2.5$\pm$92 & 75$\pm$41 \\
$\Delta_{7/2}-\Delta_{1/2}-3(N_{5/2}-N_{3/2})=0$ & 129$\pm$175 & 13$\pm$192 & 133$\pm$74 \\
 $\frac{8}{15}(\Lambda_{3/2}-N_{3/2})+\frac{22}{15}(\Lambda_{5/2}-N_{5/2})$\\$-(\Sigma_{5/2}-\Lambda_{5/2})-2(\Sigma_{7/2}''-\Delta_{7/2})=0$ & 
 91$\pm$100 & 29$\pm$75 & 0  \\
$\Lambda_{5/2}-\Lambda_{3/2}+3(\Sigma_{5/2}-\Sigma_{3/2})-4(N_{5/2}-N_{3/2})=0$ & 10$\pm$207 & 10$\pm$272 & 0  \\
$\Lambda_{5/2}-\Lambda_{3/2}+\Sigma_{5/2}-\Sigma_{3/2}-2(\Sigma_{5/2}''-\Sigma_{3/2}'')=0$ & 111$\pm$81 & 12$\pm$72 & 87$\pm$59 \\
$7(\Sigma_{3/2}''-\Sigma_{7/2}'')-12(\Sigma_{5/2}''-\Sigma_{7/2}'')=0$ & 44$\pm$319 & 39$\pm$268 & 67$\pm$266 \\
$4(\Sigma_{1/2}''-\Sigma_{7/2}'')-5(\Sigma_{3/2}''-\Sigma_{7/2}'')=0$ & 83$\pm$170 & 87$\pm$104 & 58$\pm$161 \\[.1cm]
\hline \hline
\end{tabular} }
\caption{Mass relations for the $[{\mathbf{56}}, 2^+]$ multiplet. The relations hold at the same orders as in the case of the ground state baryons. } 
\label{MassRel3}
\end{table*}

The fit and the mass relations predictions for the  experimentally unknown or poorly known states are shown in Table \ref{Predictions2}.  The PDG candidate state $\Sigma(1840)(3/2^+)^*$  could be identified with the $\Sigma_{3/2}(1889)$ state in Table \ref{Predictions2},  but as discussed earlier it can also be identified with the   Roper $\Sigma_{3/2}$. The  PDG candidate state $\Xi(2120)^*(?^?)$ is consistent with both $\Xi_{3/2}$ and $\Xi_{7/2}''$ in Table \ref{Predictions2}, so its   parity  could  be predicted as positive.

  \begin{table}[h]
  \begin{tabular}{l c c c }
  \hline
  \hline Missing states& ~Fitted mass [MeV]~ & Mass listed in PDG [MeV] & Mass  from mass relations [MeV] \\ 
     \hline $\Sigma_{3/2}$ & 1889 & $\Sigma(1840)(3/2^+)^*~\text{with mass}\sim\,{1840}$  & 1920$\pm$70 \\ 
   $\Xi_{3/2}$ & 2074 &   $\Xi(2120)^*(?^?)$:  {  2130$\pm$7} & 2080$\pm$75 \\ 
   $\Xi_{5/2}$ & 2000 & $\Xi(2030)^{***}(${\it S}$   \ \geqslant 5/2^+)$ with 2025$\pm$5 & 2006$\pm$14 \\ 
   $\Sigma_{1/2}''$ & 2059.5 & $\cdots $ & 2127$\pm$120 \\ 
   $\Xi_{1/2}''$ & 2221 & $\Xi(2250)^{**}(?^?)$:  2214$\pm$5 & $\cdots $  \\ 
   $\Omega_{1/2}$ & 2382 & $\cdots $ & $\cdots $ \\ 
   $\Sigma_{3/2}''$ & 2059.35 & $\Sigma(2080)^{**}(3/2^+)$: 2120$\pm$40 & 2109$\pm$96 \\ 
   $\Xi_{3/2}''$ & 2211.8 & $\cdots $ & $\cdots $ \\ 
   $\Omega_{3/2}$ & 2350 & $\cdots $ & $\cdots $ \\ 
   $\Sigma_{5/2}''$ & 2053 & $\Sigma(2070)^*(5/2^+)$:  2070$\pm$10 & 2077$\pm$56 \\ 
   $\Xi_{5/2}''$ & 2178 & $\cdots $ & $\cdots $ \\ 
   $\Omega_{5/2}$ & 2297 & $\cdots $ & $\cdots $ \\ 
   $\Xi_{7/2}''$ & 2129 & $\Xi(2120)^*(?^?)$: {  2130$\pm$7} & $\cdots $ \\ 
   $\Omega_{7/2}$ & 2222 & $\cdots $ & $\cdots $ \\ 
  \hline 
  \hline
  \end{tabular} 
  \caption{Predictions of physically unknown states in the $[56,2^+]$ multiplet, and suggested identifications with PDG listed states. The   first two GMO relations and the 12$^{th}$ equation in Table  \ref{MassRel3}, which is a large $N_c$ parameter independent mass relation, were used to predict the above masses.} 
  \label{Predictions2}
  \end{table}
  
\newpage

\subsection{$[{\mathbf{70}}, 1^-]$ Baryons}
 
In the case of two flavors,  there are two mixing angles  for the pairs of nucleon states with $S=1/2$ and $S=3/2$.  Denoting by $^{(2s+1)}N_S$ the nucleon state with spin $S$ and quark spin $s$, the physical states are given by:

\beq
\begin{pmatrix}
N_S \\ N'_S
\end{pmatrix}=\begin{pmatrix}
\cos \theta_{2S} & \sin \theta_{2S }\\ - \sin \theta_{2S} & \cos \theta_{2S}
\end{pmatrix} \begin{pmatrix}
^2N_{S} \\ ^4N_{S}
\end{pmatrix}.
\eeq
Understanding these mixings is very important, as the decays and photo-couplings are sensitive to them. These mixings are predicted at the leading level of breaking of spin-flavor symmetry \cite{Pirjol:2003ye}. Indeed, if the $\ord{N_c^0}$ spin-orbit operators $O_{2,3,4 }$ would have contributions of natural size, the mixing angles would be $\theta_1=\cos^{-1}(-\sqrt{2/3}\,)=2.526$ and  $ \theta_3=\cos^{-1}(-{\sqrt{5/6}}\,)=2.721$ up to $1/N_c$ corrections. However, it is known phenomenologically that the contributions of those operators are  weak, and thus the mixing angles are significantly affected by the subleading in $1/N_c$ operators, in particular the hyperfine operator $O_6$.  The determination of the mixing angles requires in principle more information than just the masses, as there are seven masses, and nine mass operators up to the order $1/N_c$, which means that the angles cannot be predicted. A biased prediction is obtained by neglecting the 3-body operators, which gives one angle as a function of the other one according the the relation \cite{deUrreta:2013koa}:
\bea
\label{eq:NLOmassrel}
&&\!\!\! 3 \left(M_{N_{\frac12}} + M_{N'_{\frac12}} - 4 M_{N_{\frac32}} - 4 M_{N'_{\frac32}} + 6 M_{N_{\frac 52}} + 8  M_{\Delta_{\frac 12}} -
    8  M_{\Delta_{\frac 32}}\right) =  \\[3.mm]
&& \left(13 \cos 2\theta_1 + \sqrt{32} \sin 2\theta_1 \right)  \left(M_{N'_{\frac12}} - M_{N_{\frac12}}\right) -
 4 \left( \cos 2\theta_3 - \sqrt{20} \sin 2\theta_3 \right) \left(M_{N'_{\frac32}} - M_{N_{\frac32}}\right) .\nonumber
  \eea
However a determination of the angles in a more rigorous way requires the input of additional observables, namely the partial decay widths and/or photo-couplings. The details of that analysis are provided in Ref. \cite{deUrreta:2013koa}.

In the case of three flavors there are two-state and also three-state mixings. For the nucleons one has the same case as for two flavors, while for $\Sigma$, $\Lambda$ and $\Xi$ baryons there is three-state mixing. The physical states are given in terms of the quark spin and $SU(3)$ eigenstates by:
\beq
\qquad \qquad \qquad \begin{pmatrix}
\bf{10}_S \text{ or } \bf{1}_S \\ 
\bf{8}_S \\ 
\bf{8'}_S
\end{pmatrix}= \Theta \begin{pmatrix}
^2\bf{10}_S \text{ or } ^2\bf{1}_S \\ 
^2\bf{8}_S \\ 
^4\bf{8}_S
\end{pmatrix},
\eeq
where the physical states are indicated by  the dominant $SU(3)$ content, and  the Euler mixing  matrix is given by:
\beq
\Theta=\begin{pmatrix} \begin{array}{ccc}
c\phi\, c\psi-\!c\theta\, s\phi\, s\psi &~  c\psi\, s\phi+c\theta\, c\phi\, s\psi &~ s\theta\, s\psi \\ 
-c\theta\, c\psi\, s\phi-\!c\phi\, s\psi &~ c\theta\, c\phi\, c\psi-\!s\phi\, s\psi & ~c\psi\, s\theta \\ 
s\theta\, s\phi &-c\phi\, s\theta & c\theta
\end{array} \end{pmatrix},
~c\theta\equiv \cos\theta,  ~s\theta\equiv \sin\theta,~ \text{etc}.
\eeq
The angles $\theta$ can always be taken in the interval $[0,\pi)$. The mixing angles $\phi$ and $\psi$ vanish in the limit of exact $SU(3)$ symmetry, and are thus proportional to the parameter $\epsilon$. The $SU(3)$ symmetric limit becomes similar to the non-strange case except that there are two additional masses, namely the ones of the singlet $\Lambda$ baryons. The determination of the mixing angles would be similar to the non-strange case.  In the absence of additional information to that of the masses, the angles can be determined only through exclusion of some operators. For instance, one strategy would be to exclude the 3-body operators, which seem   in general to have  particularly weak contributions to masses.

For the states which are subjected to mixing it is necessary to make the identification of the physical states.   As mentioned in the introduction,  for the physical case the identification has been clear for a long time, thanks to the simultaneous use of strong decay partial widths and helicity amplitudes \cite{Isgur:1978xj,Goity:2002pu,Jayalath:2011uc,deUrreta:2013koa}, but  that information is not available for the LQCD baryons. The identifications of the LQCD states were analyzed separately (a total of 256 possibilities). It turns out that most assignments pass the tests of  $\chi^2$, mass relations and naturalness of the coefficients. Thus on a general rigorous ground the problem of state assignment is not completely resolved. However, if one requires that the coefficients flow reasonably smoothly towards the physical ones which are known, then only one assignment becomes possible, namely the one discussed here.

Since the mixing angles would be an indicator of level-crossing effects as $M_\pi$ is varied, their definite understanding is an important task. In fact,  recent studies of lower lying negative parity states in Ref. \cite{Engel:2013ig,Mahbub:2013ala} identified the two lowest lying $N_{1/2}^-$ masses  and may give  the first evidence of such a level crossing as $M_\pi$ varies.

For two flavors, and ollowing the global analysis of Ref \cite{deUrreta:2013koa}, the two mixing angles are given as input, and in this way it is possible to fit with the complete basis of operators up to 2-body. If the additional information provided by partial decay widths and/or photocouplings is not available, as it is the case for the LQCD results, one possibility is to neglect some of the basis operators, which allows one to predict the mixing angles solely using the masses. A guidance on what operator(s) to exclude is given by the rather clear hierarchy in the importance the different operators have, as measured by the magnitude of their coefficients. In fact, it becomes clear that the mixing angles are mostly controlled by the operators $O_2$, $O_6$ and to a lesser extend $O_4$ and $O_5$. 
In the case of three flavors the number of masses is much larger than the number of basis operators, and thus in principle the mixing angles can be determined with the information on the masses, of course after the above mentioned identification of states has been performed.  

\begin{figure}[h]\vspace*{5mm}~\\
\centering
\includegraphics[width=0.45\linewidth]{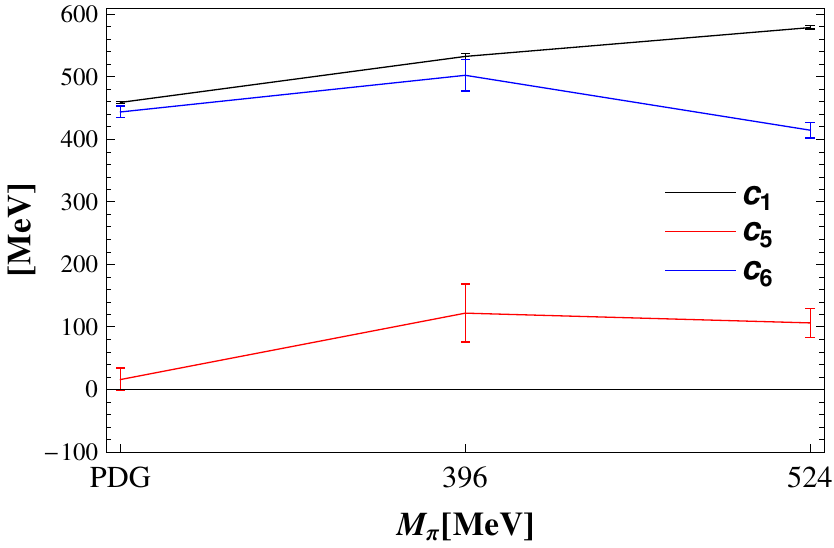}
\caption{Evolution of the minimum set of operator coefficients with $M_\pi$ in $SU(4)\times O(3)$}
\label{fig:EvolutiontestPic1}
\end{figure}

For  two flavors,  the LQCD results are those in Ref. \cite{Edwards:2011jj}, and the corresponding  fits are shown in Table \ref{Fit201minus} \footnote{In order to compare with the coefficients $C_i$ obtained in the global analysis \cite{deUrreta:2013koa}, where the operators are given in spherical basis and  with  different normalizations than here, the correspondence is: $C_1=c_1$,  $C_2=-\frac{5}{6}c_2$,  
$C_3=-\frac{75}{144} c_3$,  $C_4=\frac{3}{2} c_4$,  $C_5=-\frac{5}{3} c_5$,  $C_6=2 c_6$,  $C_7=-c_7$, and $C_8=\frac 53 c_8$.
}.
 The physical case is in good agreement with previous works \cite{Carlson:1998gw,Carlson:1998vx}.  If one considers only the seven known masses as inputs to the fit, one operator must be eliminated: the operator $O_8$ is thus dismissed as it always results virtually irrelevant.  A second fit where only the three dominant operators are kept turns out to be  consistent for the lattice QCD results, but gives a poor fit to the physical case. In that case, the  $M_\pi$ evolution  of the coefficients is shown in Fig. \ref{fig:EvolutiontestPic1}.

\begin{table}
\setlength{\tabcolsep}{6pt}
\begin{tabular}{l c c c c c c}
\hline
\hline
Coefficients & & & $M_\pi$[MeV]& & & \\ 
$\text{[MeV]}$  & PDG & 396 & 524 & PDG & 396 & 524\\ \hline
 $c_1$   & 463$\pm$2 & 543$\pm$5 & 598$\pm$3 & 459$\pm$2 & 533$\pm$5 & 579$\pm$3 \\ 
 $c_2$   & -36$\pm$12 & 39$\pm$35 & 13$\pm$14 & 0 & 0 & 0 \\ 
 $c_3$   & 313$\pm$69 & -83$\pm$215 & -96$\pm$74 & 0 & 0 & 0 \\  
 $c_4$   & 65$\pm$31 & -70$\pm$71 & -95$\pm$30 & 0 & 0 & 0 \\  
 $c_5$   & 71$\pm$18 & 99$\pm$48 & 107$\pm$24 & 16$\pm$18 & 122$\pm$46 & 106$\pm$23 \\  
 $c_6$   & 443$\pm$10 & 446$\pm$25 & 307$\pm$13 & 443$\pm$10 & 502$\pm$25 & 414$\pm$13\\  
 $c_7$   & -20$\pm$31 & -0.37$\pm$62.89 & -66$\pm$34 & 0 & 0 & 0 \\  
 \hline
 $\theta_{N_{1/2}}$ $[\text{Rad}]$& 0.52$\pm$0.13 & 2.94$\pm$0.21 & 2.76$\pm$0.06 & 3.13$\pm$0.01 & 3.04$\pm$0.05 & 3.03$\pm$0.03 \\ 
  $\theta_{N_{3/2}}$ $[\text{Rad}]$ & 3.02$\pm$0.09 & 2.88$\pm$0.42 & 2.38$\pm$0.11 & 3.12$\pm$0.02 & 2.98$\pm$0.08 & 2.97$\pm$0.05 \\ 
  \hline {\small$\chi^2_{\rm dof}$}~~~ & 0.05 & 0  & 0  & 0.68 & 0.52 & 1.0\\ 
\hline 
\hline
\end{tabular}  
\caption{Fits to the non-strange $[{\bf{20}},1^-]$ baryon masses. Unless the mixing angles are inputs to the fit, the operator $O_8$ is not necessary due to linear dependence as there are only seven mass inputs to fit.   For the physical case with seven parameter fit, the mixing angles from the global analysis ($\theta_{N_{1/2}}$=0.49$\pm$0.29, $\theta_{N_{3/2}}$=3.01$\pm$0.17) were used as inputs.  For the minimal fit with $c_1$,$c_5$,$c_6$, the mixing angles in the physical case are not inputs.} 
\label{Fit201minus}
\end{table}


A comparison of the physical case shows that it is consistent with earlier work \cite{Carlson:1998gw,Carlson:1998vx}, but differs  significantly  for the coefficients $c_3$ and $c_6$   with respect to  the  recent global analysis carried out in Ref. \cite{deUrreta:2013koa}.   Since all those  fits are consistent in terms of the $\chi^2$, it is indication of the ambiguity   that  results  when only the masses are fitted. This means that also for the LQCD results one should expect several consistent fits in terms of the value of the $\chi^2$,  which will have some of the parameters significantly different.  


 Now the fits to the three flavor case are presented. The identification of the states has been made as described earlier.
 Such identification is clearly  displayed in Table \ref{Masses701minus}  of Appendix B.  For the sake of brevity, only those operators which have effects of any significance have been included here: after an initial analysis, several operators whose coefficients resulted consistent with zero have been eliminated.   The fits for three flavors are given in Tables \ref{Fit701minus1} and \ref{Fit701minus2} for  the corresponding subsets of operators.
 Because of the different definitions of the basis operators for the different multiplets, in order to compare contributions which are of common nature across mutliplets such as the spin-flavor singlet contributions, the HF and the $SU(3)$ breaking, the following identification of coefficients should be done: $c_{1_{\mathbf{56}}}\leftrightarrow (c_1+(b_1+b_2)/\sqrt{3})_{\mathbf{70}}$, $ c_{2_{[\mathbf{56},0^+]}}\leftrightarrow  c_{3_{[\mathbf{56},2^+]}}\leftrightarrow  \frac{1}{3}c_{6_{\mathbf{70}}} $,  $b_{1_{\mathbf{56}}}\leftrightarrow -\left((b_1+b_2)\sqrt{3}/2\right)_{\mathbf{70}}$.

 \begin{table*}[htpb!!]
  \setlength{\tabcolsep}{7pt}
  \begin{tabular}{ l c c c c }
  \hline 
  \hline
  Coefficients & & & $M_\pi$[MeV]&  \\
  $[\text{MeV}]$  & PDG & 391 & 524 & 702 \\ 
  \hline 
   $c_1$   & 444.3$\pm$0.3 & 572$\pm$2 & 585$\pm$1 & 636$\pm$1 \\ 
   $c_2$   & 84$\pm$2 & 68$\pm$12 & -7$\pm$6 & -16$\pm$4  \\ 
   $c_3$   & 117$\pm$13 & 59$\pm$22 & -40$\pm$18 & 2$\pm$8 \\ 
   $c_4$   & 115$\pm$5 & -12$\pm$12 & -28$\pm$9 & -13$\pm$4 \\ 
   $c_5$   & 84$\pm$10 & 134$\pm$17 & 132$\pm$14 & 84$\pm$7  \\ 
   $c_6$   & 538$\pm$5 & 327$\pm$10 & 350$\pm$6 & 262$\pm$4 \\ 
   $c_7$   & -159$\pm$13 & 49$\pm$27 & -59$\pm$17 & 13$\pm$11 \\ 
   $b_1$   & -214$\pm$5 & -100$\pm$13 & -43$\pm$9 & 0 \\ 
   $b_2$   & -188$\pm$2 & -62$\pm$6 & -46$\pm$4 & 0 \\ 
   $b_3$   & -92$\pm$2 & -41$\pm$10 & -6$\pm$7 & 0 \\ 
   \hline $\chi^2_{\rm dof}$ & 0.74 & 0.65 & 0.14 & 0.09\\ 
   \hline\hline
  \end{tabular} 
  \caption{{Fit to the $[{\bf{70}},1^-]$  masses using a subset of operators chosen as a minimal subset such the $\chi^2_{\rm dof}$ is acceptable for all input sets.  For the physical case the mixing angles from the global analysis \cite{deUrreta:2013koa} ($\theta_{N_{1/2}}$=0.49$\pm$0.29, $\theta_{N_{3/2}}$=3.01$\pm$0.17) were used as inputs.} }
  \label{Fit701minus1}
  \end{table*}

 \begin{table}
 \setlength{\tabcolsep}{7pt}
 \begin{tabular}{ l c c c c}
 \hline 
 \hline
  Mixing angles   & & & $M_\pi$[MeV]&  \\
  $[\text{Rad}]$ & PDG & 391 & 524 & 702 \\ 
   \hline
  $\theta_{N_{1/2}}$ & 0.76$\pm$0.03 & 0.61$\pm$0.12 & 2.77$\pm$0.06 & 2.98$\pm$0.05  \\ 
  $\theta_{N_{3/2}}$ & 3.09$\pm$0.40 & 0.10$\pm$0.81 & 2.70$\pm$0.10 & 2.84$\pm$0.03 \\ 
  \hline
  $\phi_{\Lambda_{1/2}}$ & -0.15$\pm$0.01  & ~~-0.15$\pm$0.01 & -0.14$\pm$0.01 & 0 \\
  $\theta_{\Lambda_{1/2}}$ & 0.83$\pm$0.01 & 0.70$\pm$0.01 & 2.76$\pm$0.01 & 2.98$\pm$0.05   \\ 
   $\psi_{\Lambda_{1/2}}$ & 0.05$\pm$0.01 & 0.11$\pm$0.01 & -0.18$\pm$0.02 & 0 \\
   \hline
  $\phi_{\Lambda_{3/2}}$ & -0.21$\pm$0.03 & -0.16$\pm$0.04 & -0.12$\pm$0.02 & 0 \\
  $\theta_{\Lambda_{3/2}}$ & 3.08$\pm$0.01 & 0.13$\pm$0.01 & 2.69$\pm$0.02 &  2.84$\pm$0.03\\
   $\psi_{\Lambda_{3/2}}$ & -0.18$\pm$0.01 & 0.07$\pm$0.03 & -0.03$\pm$0.01 & 0 \\
   \hline
    $\phi_{\Sigma_{1/2}}$ & -0.25$\pm$0.02 & 0.03$\pm$0.01 & -0.05$\pm$0.04 & 0 \\
   $\theta_{\Sigma_{1/2}}$ & 1.01$\pm$0.01 & 0.75$\pm$0.01 & 2.75$\pm$0.01 & 2.98$\pm$0.05   \\
    $\psi_{\Sigma_{1/2}}$ & -0.10$\pm$0.01 & 0.01$\pm$0.07 & 0.03$\pm$0.04 & 0 \\
  \hline
      $\phi_{\Sigma_{3/2}}$ & -0.08$\pm$0.06 & 0.06$\pm$0.04 & -0.02$\pm$0.04 & 0 \\
      $\theta_{\Sigma_{3/2}}$ & 3.05$\pm$0.01 & 0.16$\pm$0.02 & 2.66$\pm$0.01  & 2.84$\pm$0.03\\
       $\psi_{\Sigma_{3/2}}$ & 0.04$\pm$0.02 & 0.03$\pm$0.02 & 0.005$\pm$0.001 & 0\\
 \hline      
      $\phi_{\Xi_{1/2}}$ & -0.30$\pm$0.03 & 0.03$\pm$0.01 & -0.05$\pm$0.06 & 0 \\
     $\theta_{\Xi_{1/2}}$ & 0.94$\pm$0.01 & 0.78$\pm$0.01 & 2.77$\pm$0.04 & 2.98$\pm$0.05   \\ 
      $\psi_{\Xi_{1/2}}$ & -0.14$\pm$0.02 & 0.01$\pm$0.07 & 0.03$\pm$0.06 & 0 \\
     \hline
     $\phi_{\Xi_{3/2}}$ & -0.09$\pm$0.07 & 0.05$\pm$0.03 & -0.02$\pm$0.04 & 0 \\
    $\theta_{\Xi_{3/2}}$ & 3.07$\pm$0.01 & 0.19$\pm$0.03 & 2.69$\pm$0.02 & 2.84$\pm$0.03\\
    $\psi_{\Xi_{3/2}}$ & 0.05$\pm$0.03 & 0.02$\pm$0.01 & 0.006$\pm$0.001 & 0 \\
 \hline \hline
 \end{tabular} 
 \caption{Mixing angles in the  $[{\bf {70}},1^-]$ predicted from the fit to the masses.   }
 \label{Mixing701minus}
 \end{table}

The fits in the physical case are checked to be consistent with previous analysis \cite{Schat:2001xr,Goity:2002pu}. It is interesting to observe the evolution of the mixing angles $\theta$ with $M_\pi$, as  they can give a clue on the possible level crossing as $M_\pi$ evolves. As it is the case in the non-strange case discussed above, in the $S=3/2$ baryons these angles remain continuous  from  the physical case to $M_\pi=702$ MeV, while in the case of the $S=1/2$ baryons there is a change by more than $\pi/2$, indicating a level crossing along the way.  This qualitatively  agrees with the LQCD results in Refs. \cite{Engel:2013ig,Mahbub:2013ala}.  It is interesting to observe that for $M_\pi=702$ MeV all baryons are stable, and almost all are still stable for $M_\pi=524$ MeV, while below $M_\pi=391$ MeV they are unstable.  Since the $S=1/2$ baryons have  S-wave decays, they are the ones to be sensitive to the opening of the decay. These observations suggest a synchronization between the mixing angle and the stability of the baryon. In fact, the change in $\theta_1$ shown in Table \ref{Mixing701minus}  in going from $M_\pi=391$ to $524$ MeV is approximately  $\pi/2$,  as expect for a level crossing. Is this an explanation for the observed  level crossings?.  Perhaps, but it is not clear at this point, and it deserves further study.

\begin{figure}[h]
\centering
\includegraphics[width=0.65\linewidth]{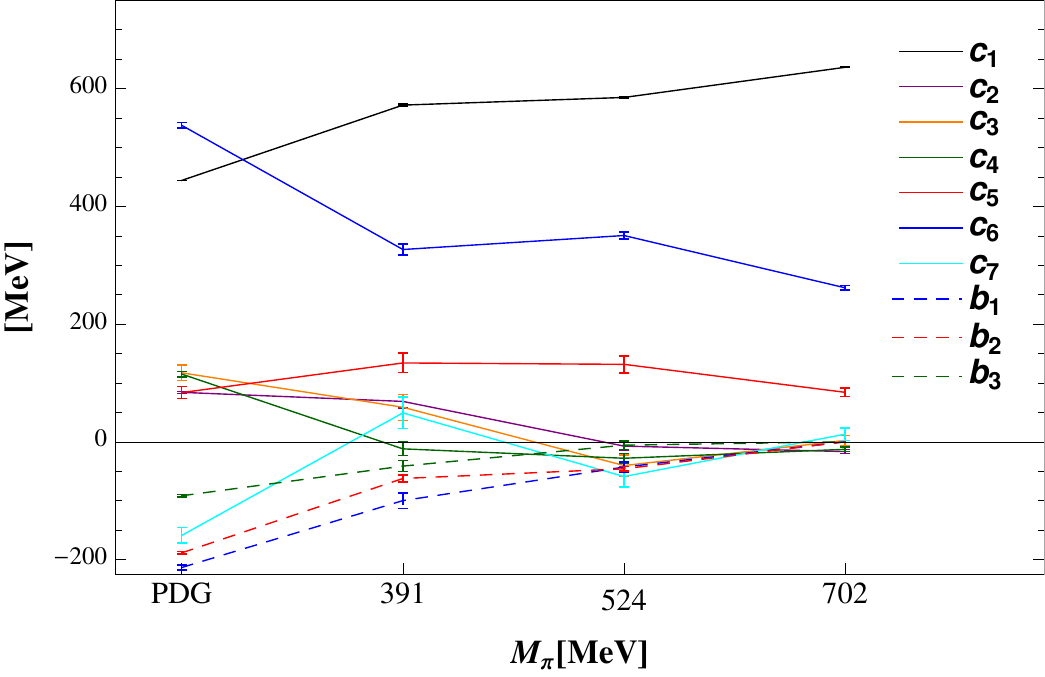}
\caption{Evolution with respect to $M_{\pi}$ of the  coefficients of the basis operators used to fit both the physical and the LQCD $[{\bf{70}},1^-]$ masses. 
}
\label{fig:UnionSetNegative70plet}
\end{figure}


 
\begin{figure}[h]
\centering
\includegraphics[width=0.52\linewidth]{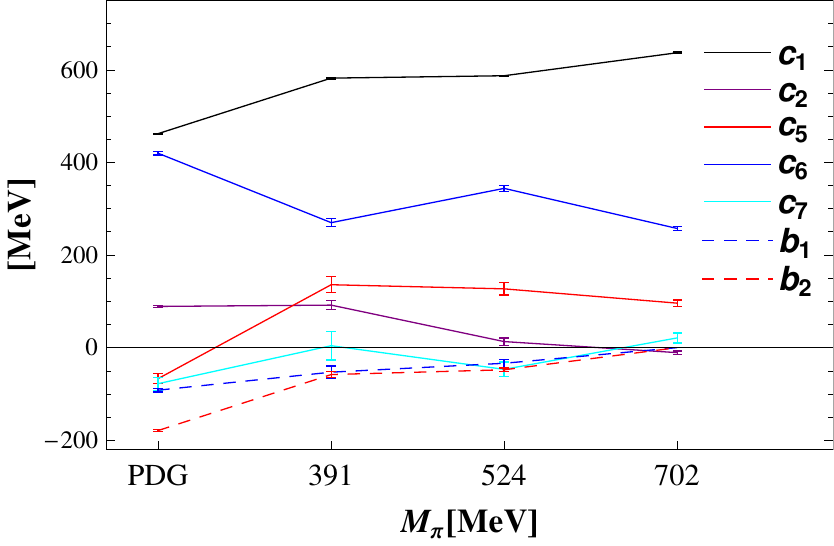}
\caption{{Evolution with respect to $M_{\pi}$ of the  coefficients  in Table \ref{Fit701minus2}. }. 
 }
\label{fig:MinimumSetNegative70plet}
\end{figure}

 Consistent fits  to  only LQCD results can be achieved by a minimal set of significant operators. It is found that the relevant $SU(3)$ singlet operators are the spin-flavor singlet $O_1$, the HF $O_6$ and the two spin-orbit ones $O_2$ and $O_5$ and the first two $SU(3)$ breaking operators. These results are illustrated in Fig. \ref{fig:MinimumSetNegative70plet}.
Note that all the $SU(3)$ breaking operators are relevant for fitting  the physical case.  The  operator $O_3$ is found to be important for the  physical masses, but irrelevant for the LQCD   masses, where the operator $O_5$  is instead significant.  It is interesting to observe that in models with pion exchange between quarks, such as certain versions of the chiral quark model, $O_3$ is naturally important, and should fade as the $M_\pi$ increases.

\begin{table*}[htpb!!]
\setlength{\tabcolsep}{7pt}
\begin{tabular}{ l c c c c }
\hline 
\hline
Coefficients & & & $M_\pi$[MeV]&  \\
$[\text{MeV}]$  & PDG & 391 & 524 & 702 \\ 
\hline 
 $c_1$   & 462$\pm$0.3 & 582$\pm$2 & 587$\pm$1 & 637$\pm$1 \\ 
 $c_2$   & 83$\pm$2 & 92$\pm$10 & 13$\pm$8 & -11$\pm$4  \\ 
  $c_5$   & -67$\pm$11 & 136$\pm$17 & 127$\pm$13 & 96$\pm$7  \\ 
 $c_6$   & 420$\pm$4 & 270$\pm$9 & 344$\pm$6 & 257$\pm$4 \\ 
 $c_7$   & -78$\pm$14 & 4$\pm$31 & -47$\pm$16 & 21$\pm$11 \\ 
 $b_1$   & -92$\pm$4 & -53$\pm$13 & -34$\pm$9 & 0 \\ 
 $b_2$   & -179$\pm$2 & -58 $\pm$6 & -48$\pm$4 & 0 \\ 
     \hline
 $\theta_{N_{1/2}}$ & 0.33$\pm$0.02 & 0.79$\pm$0.21 & 2.95$\pm$0.05 & 2.94$\pm$0.02  \\ 
 $\theta_{N_{3/2}}$ & 0.45$\pm$0.02 & 0.79$\pm$0.13 & 2.86$\pm$0.07 & 2.84$\pm$0.03 \\ 
 \hline $\chi^2_{\rm dof}$ & 6.7 & 0.86 & 0.46 & 0.13\\ 
   \hline 
\hline 
\end{tabular} 
\caption{  Fit results with minimal set of mass operators for the $[{\bf{70}},1^-]$.  Only masses are used as inputs.}
\label{Fit701minus2}
\end{table*}

 The mass relations are depicted in Tables \ref{MassRel4} and \ref{MassRel5}. All are well satisfied, except for the EQS relation for $M_\pi=391$ MeV involving $\Sigma^{''}_{3/2}$. A shift of its mass by $\sim +30 $ MeV leads to consistency.   The mass predictions are given in Table \ref{predictions31}. Since, the PDG candidate state $\Xi(1950)^{***}(?^?)$ is consistent with $\Xi_{3/2}'$,$\Xi_{5/2}$ and $\Xi_{1/2}''$ in Table \ref{predictions31},   its parity could be predicted as negative. 
\begin{table*}[htpb!]
\setlength{\tabcolsep}{6pt}
\begin{tabular}{l c c c}
\hline
\hline Relation & & $M_\pi$[MeV] & \\
  & PDG & 391 & 524\\ 
\hline
$2(N_{1/2}+\Xi_{1/2})-(3\Lambda_{1/2}+\Sigma_{1/2})=0$ & $\cdots $ & 59$\pm$156 & 17$\pm$125  \\
$2(N_{3/2}+\Xi_{3/2})-(3\Lambda_{3/2}+\Sigma_{3/2})=0$ & $\cdots $ & 31$\pm$121 & 13$\pm$74 \\
$2(N_{5/2}+\Xi_{5/2})-(3\Lambda_{5/2}+\Sigma_{5/2})=0$ & $\cdots $ & 46$\pm$91 & 6$\pm$64 \\
$\Sigma_{1/2}''-\Delta_{1/2}=\Xi_{1/2}''-\Sigma_{1/2}''=\Omega_{1/2}-\Xi_{1/2}''$ & $\cdots $ & 67$\pm$47 & 35$\pm$56  \\
 & $\cdots $ & 34$\pm$36 & 40$\pm$41  \\
 & $\cdots $ & 24$\pm$49 & 22$\pm$26  \\
$\Sigma_{3/2}''-\Delta_{3/2}=\Xi_{3/2}''-\Sigma_{3/2}''=\Omega_{3/2}-\Xi_{3/2}''$ & $\cdots $ & 2$\pm$49 & 39$\pm$23 \\
 & $\cdots $ & 82$\pm$47 & 37$\pm$21  \\
 & $\cdots $ & 61$\pm$43 & 31$\pm$21  \\
\hline \hline
\end{tabular} 
\caption{GM-O and ES relations for the $[{\mathbf{70}}, 1^-]$ multiplet.  Due to the insufficient number of physically known   states with three or more stars, the mass relations for physical states cannot be checked for the physical case.}
\label{MassRel4}
\end{table*}




\begin{table*}[h]
\begin{tabular}{l c c }
\hline
\hline Relation &   ~~~~~~~~~~~~~~~~~~$~M_\pi$\,[MeV]&   \\
  & 391 & 524   \\ 
\hline
$14(S_{\Lambda_{3/2}}+ S_{\Lambda_{3/2}'})+63S_{\Lambda_{5/2}}+ 36(S_{\Sigma_{1/2}}+ S_{\Sigma_{1/2}'})$\\$ -68(S_{\Lambda_{1/2}}+ S_{\Lambda_{1/2}'})- 27S_{\Sigma_{5/2}}=0$ & 9.4$\pm$40 & 0.96$\pm$34   \\
$14(S_{\Sigma_{3/2}}+ S_{\Sigma_{3/2}'})+21S_{\Lambda_{5/2}}- 9S_{\Sigma_{5/2}}$\\$-18(S_{\Lambda_{1/2}}+ S_{\Lambda_{1/2}'})- 2(S_{\Sigma_{1/2}}+ S_{\Sigma_{1/2}'})=0$ & 37$\pm$45  & 5.4$\pm$38 \\
$14\,S_{\Sigma_{1/2}''}+49S_{\Lambda_{5/2}}+ 23(S_{\Sigma_{1/2}}+ S_{\Sigma_{1/2}'})$\\$-45(S_{\Lambda_{1/2}}+ S_{\Lambda_{1/2}'})- 19S_{\Sigma_{5/2}}=0$ & 9.4$\pm$40 & 0.7$\pm$34   \\ 
$14\,S_{\Sigma_{3/2}''}+28S_{\Lambda_{5/2}}+ 11(S_{\Sigma_{1/2}}+ S_{\Sigma_{1/2}'})$\\$-27(S_{\Lambda_{1/2}}+ S_{\Lambda_{1/2}'})- 10S_{\Sigma_{5/2}}=0$  & 0.8$\pm$40 & 0.1$\pm$33   \\ [.1cm]
\hline \hline
\end{tabular} 
\caption{Octet-Decuplet mass relations for the 
$[{\mathbf{70}}, 1^-]$ multiplet. 
 $S_{B}$ is the mass splitting between the state $B$ and the non-strange states in the $SU(3)$ multiplet to which it belongs.
The results  shown  correspond to the relation   divided by the sum of the positive coefficients in the relation (e.g., 163 for the first relation).
 }
 \label{MassRel5}
\end{table*}
%
\begin{table}[H]
\begin{tabular}{ c c c }
\hline
\hline ~~~Missing State~~ & ~~~Fitted mass with union set of operators & ~~PDG \\ 
  & [MeV] & ~~[MeV] \\
\hline $\Sigma_{1/2}$ & 1644.72 & $\Sigma(1620)1/2^{-**}$=1620$\pm$10 \\ 
 $\Xi_{1/2}$ & 1800.93 & $\cdots $ \\ 
 $\Xi_{1/2}'$ & 1930.24 & $\cdots $ \\ 
 $\Lambda_{3/2}'$ & 1824.59 & $\cdots $ \\ 
 $\Sigma_{3/2}'$ & 1780.37 & $\cdots $ \\ 
 $\Xi_{3/2}'$ &  1943.64 & $\Xi(1950)(?^?)^{***}$=1950$\pm$15 \\ 
 $\Xi_{5/2}$ & 1938.95 & $\Xi(1950)(?^?)^{***}$=1950$\pm$15 \\ 
 $\Sigma_{1/2}''$ & 1827.51 & $\cdots $ \\ 
 $\Xi_{1/2}''$ & 1968.76 & $\Xi(1950)(?^?)^{***}$=1950$\pm$15 \\ 
 $\Omega_{1/2}$ & 2107.31 & $\cdots $ \\ 
 $\Sigma_{3/2}''$ & 1916.21 & $\Sigma(1940)3/2^{-***}$=1950$\pm$30 \\ 
 $\Xi_{3/2}''$ & 2057.24 & $\cdots $ \\ 
 $\Omega_{3/2}$ & 2197.75 & $\cdots $ \\ 
\hline 
\hline
\end{tabular} 
\caption{Predictions of physically unknown states in the $[70,1^-]$ multiplet from the fit in Table \ref{Fit701minus1}.
}
\label{predictions31} 
\end{table}


\section{Comments and conclusions}

From the study presented here of recent LQCD results for the low lying baryon excitations, it can be concluded that a clear picture of  their  spin-flavor composition can be obtained. This entirely  supports  the  picture  seen from the lattice QCD analysis of the mass eigenstate couplings to source/sink operators.  A similar, and even simpler picture than   the physical case emerges at increasing quark masses, where with very few dominant operators the LQCD masses can be described. The expected narrowness of the states analyzed for the quark masses in the LQCD results suggests that those results are very realistic. For higher excited baryons, which will be broader, the present LQCD results may be a poorer approximation. Nonetheless, they should  be interesting to study.

 A strong conclusion is that the LQCD masses are even closer to an approximate $SU(6)\times O(3)$ symmetry limit than the physical ones. This is most likely due to the fact that the composition of baryons becomes increasingly closer to a constituent quark model picture as the quark masses increase, emphasizing the mass operators which are naturally large in  those models and suppressing the rest.
 The study presented here shows that the LQCD masses  can in all cases be described quite well with only a few operators, namely the leading spin-flavor singlet one, the hyperfine one and with a lesser relevance the spin-orbit one.  
 
For the  quark masses employed in the LQCD calculations used here,  the dramatic downturn in $c_1$ for the Roper baryons is not manifest.  This is an effect where probably chiral symmetry  plays an important role, but it is not evident. In recent LQCD work on nucleon resonances  \cite{Engel:2013ig,Mahbub:2013ala} a first evidence of  that  downturn is observed. It remains to figure out what is the precise mechanism that drives that effect, perhaps using some clever strategy in the LQCD calculation. While in the Roper multiplet the $c_1$ coefficient should have that large negative curvature as a function of $M_\pi$ to match the physical masses, it lies along an almost prefect straight line for the ground state baryons, and it has a moderate negative curvature in the other cases.

Identifying the HF coefficients as mentioned earlier, one finds that for the LQCD results the strength of the HF in the ground state baryons is almost twice as large as in the excited baryons, which is significantly different than   the physical case, where it is only about 25\% larger. 

An interesting open problem is how to relate the $SU(6)\times O(3)$ decomposition of the physical baryons determined via the $1/N_c$ expansion as presented here,  with the information on the coupling strengths of the mass eigenstates to the different source/sink operators obtained in the LQCD calculations. 

\begin{acknowledgments}
 
 The authors thank Jozef Dudek  and  Robert Edwards for   useful discussions.
This work was supported in part  %
by DOE Contract No. DE-AC05-06OR23177 under which JSA operates the Thomas Jefferson National Accelerator Facility (J.~L.~G.), and by the National Science Foundation  through grants  PHY-0855789 and PHY-1307413 (I.~P.~F. and J.~L.~G.). 
\end{acknowledgments}

\newpage

\appendix

\section{ Bases of mass operators}

This appendix gives the bases of mass operators with the respective normalization factors used in this work.

\subsection{Operator basis and matrix elements for the  $[{\mathbf{56}}, 2^+]$ multiplet.}

\begin{table}[h!]
\setlength{\tabcolsep}{8pt}
\renewcommand{\arraystretch}{.85}
{\squeezetable
\begin{tabular}{ c c c c }
\hline\hline  & $O_1$ & $O_2$ & $O_3$ \\ 
 &  $N_c\bf{1}$ & $\frac{1}{N_c}\ell_i\,s_i$ & $\frac{1}{N_c}S_iS_i$\\[.2cm]
\hline $8_{3/2}$ & $N_c$ & $-\frac{3}{2N_c}$ & $\frac{3}{4N_c}$ \\ 
$8_{5/2}$ & $N_c$ & $\frac{1}{N_c}$ & $\frac{3}{4N_c}$ \\ 
 $10_{1/2}$ & $N_c$ & $-\frac{9}{2N_c}$ & $\frac{15}{4N_c}$ \\ 
 $10_{3/2}$ & $N_c$ & $-\frac{3}{N_c}$ & $\frac{15}{4N_c}$ \\ 
 $10_{5/2}$ & $N_c$ & $-\frac{1}{2N_c}$ & $\frac{15}{4N_c}$ \\ 
 $10_{7/2}$ & $N_c$ & $\frac{3}{N_c}$ & $\frac{15}{4N_c}$ \\ [.2cm]
\hline 
\hline
& $\bar{{B_1}}$ & $\bar{{B_2}}$ & $\bar{{B_3}}$ \\ 
 & $N_s$ & $\small \frac{1}{N_c}\ell_{i}G_{i8}-\frac{1}{2\sqrt{3}}O_2$ & $ \small \frac{1}{N_c}S_{i}G_{i8}-\frac{1}{2\sqrt{3}}O_3$\\[.2cm]
\hline $N_S$ & 0 & 0 & 0 \\ 
 $\Lambda_S$ & 1 & $\frac{3\sqrt{3}\;a_S}{4N_c}$ & $-\frac{3\sqrt{3}}{8N_c}$ \\ 
 $\Sigma_S$ & 1 & $-\frac{\sqrt{3}\;a_S}{4N_c}$ & $\frac{\sqrt{3}}{8N_c}$ \\ 
 $\Xi_S$ & 2 & $\frac{\sqrt{3}\;a_S}{N_c}$ & $-\frac{\sqrt{3}}{2N_c}$ \\ 
 $\Delta_S$ & 0 & 0 & 0 \\ 
 $\Sigma_S''$ & 1 & $\frac{3\sqrt{3}\;b_S}{4N_c}$ & $-\frac{5\sqrt{3}}{8N_c}$ \\ 
 $\Xi_S''$ & 2 & $\frac{3\sqrt{3}\;b_S}{2N_c}$ & $-\frac{5\sqrt{3}}{4N_c}$ \\ 
 $\Omega_S$ & 3 & $\frac{9\sqrt{3}\;b_S}{4N_c}$ & $-\frac{15\sqrt{3}}{8N_c}$ \\ 
 $\Sigma_{3/2}-\Sigma_{3/2}''$ & 0 & $\frac{\sqrt{3}}{2N_c}$ & 0 \\ 
 $\Sigma_{5/2}-\Sigma_{5/2}''$ & 0 & $\frac{\sqrt{3}}{2N_c}$ & 0 \\ 
 $\Xi_{3/2}-\Xi_{3/2}''$ & 0 & $\frac{\sqrt{42}}{6N_c}$ & 0 \\ 
 $\Xi_{5/2}-\Xi_{5/2}''$ & 0 & $\frac{\sqrt{42}}{6N_c}$ & 0 \\ [.2cm]
\hline \hline
\end{tabular} 
\caption{Matrix elements of $SU(3)$ singlet operators (top)  and $SU(3)$ breaking operators (bottom). Here, $a_S=1,\;-2/3$ for $S=3/2,\;5/2$, respectively and $b_S=1,\;2/3,\;1/9,\;-2/3$ for $S=1/2,\;3/2,\;5/2,\;7/2$,  respectively}
\label{562plusbasis}}
\end{table}
\newpage 
 
 \subsection{Operator basis and matrix elements for the  $[\mathbf{20}, 1^-]$ multiplet.}

\begin{table}[h!]
\setlength{\tabcolsep}{8pt}
\renewcommand{\arraystretch}{.85}
{\squeezetable
\setlength{\tabcolsep}{4pt}
  \begin{tabular}{c c c c c  }
  \hline\hline
  & $O_1$ & $O_2$ & $O_3$ & $O_4$ 
   \\ 
   & $N_c\bf{1}$ & $\ell_i\,s_i$ & $\frac{3}{N_c}\ell_{ij}^{(2)}g_{ia}G_{ja}^c$ & $\ell_i\,s_i+ \frac{4}{N_c+1}\,\ell_i\,t_aG_{ja}^c$ 
    \\ [.2cm]
  \hline 
  $N_{1/2}$ &
   $N_c$ & $- \frac{(2N_c-3)}{3N_c}$ & 0 & $ \frac{2}{N_c+1}$ 
    \\ 
   $N_{1/2}'$ &
    $N_c$ & - $\frac{5}{6}$ & $- \frac{5(N_c+1)}{48N_c}$ & 0 
     \\ 
   $N_{1/2}'- N_{1/2}$ & 
   0 & $-\frac{1}{3} \sqrt{\frac{N_c+3}{2N_c}}$ & $-\frac{5}{48N_c}\sqrt{\frac{(N_c+3)(2N_c-1)^2}{2N_c}}$ & $- \sqrt{\frac{N_c+3}{2N_c(N_c+1)^2}}$
    \\ 
  $N_{3/2}$ &
   $N_c$ & $ \frac{(2N_c-3)}{6N_c}$ & 0 & $- \frac{1}{N_c+1}$ 
   \\ 
  $N_{3/2}'$  &
   $N_c$ & $- \frac{1}{3}$ & $ \frac{1}{12N_c}(N_c+1)$ & 0 
   \\ 
   $N_{3/2}'- N_{3/2}$& 
   0 & $-\frac{1}{6} \sqrt{\frac{5(N_c+3)}{N_c}}$ & $\frac{1}{96N_c} \sqrt{\frac{5(N_c+3)(2N_c-1)^2}{N_c}}$ & $- \sqrt{\frac{5(N_c+3)}{4N_c(N_c+1)^2}}$ 
   \\ 
   $N_{5/2}'$ &
    $N_c$ & $+\frac{1}{2}$ & $-\frac{1}{48N_c}(N_c+1)$ & 0
    \\ 
  $\Delta_{1/2}$
   &
    $N_c$ & $+\frac{1}{3}$ & 0 & 0
    \\ 
  $\Delta_{3/2}$ 
  & 
  $N_c$ & $-\frac{1}{6}$ & 0 & 0 
   \\ [.2cm]
  \hline 
  \hline
  & $O_5$ & $O_6$ & $O_7$ & $O_8$ 
   \\ 
 & $\frac{1}{N_c}\,\ell_i\,S_i^c$ & $\frac{1}{N_c}S_i^cS_i^c$ & $\frac{1}{N_c}s_iS_i^c$ & $\frac{2}{N_c}\ell_{ij}^{(2)}s_iS_j^c$ 
    \\ [.2cm]
  \hline 
  $N_{1/2}$ 
   & $- \frac{(N_c+3)}{3N_c^2}$ & $ \frac{(N_c+3)}{2N_c^2}$ & $- \frac{(N_c+3)}{4N_c^2}$ & 0 
    \\ 
   $N_{1/2}'$ 
  & $- \frac{5}{3N_c}$ & $ \frac{2}{N_c}$ & $ \frac{1}{2N_c}$ & $ \frac{5}{6N_c}$ 
     \\ 
   $N_{1/2}'- N_{1/2}$ 
    & $\sqrt{\frac{N_c+3}{18N_c^2}}$ & 0 & 0 & $\frac{5}{12N_c} \sqrt{\frac{N_c+3}{2N_c}}$ 
    \\ 
  $N_{3/2}$ 
 & $ \frac{(N_c+3)}{6N_c^2}$ & $ \frac{(N_c+3)}{2N_c^2}$ & $- \frac{(N_c+3)}{4N_c^2}$ & 0 
   \\ 
  $N_{3/2}'$  
 & $-\frac{2}{3N_c}$ & $\frac{2}{N_c}$ & $\frac{1}{2N_c}$ & $-\frac{2}{3N_c}$ 
   \\ 
   $N_{3/2}'- N_{3/2}$
 & $ \sqrt{\frac{5(N_c+3)}{36N_c^3}}$ & 0 & 0 & $-\frac{1}{24} \sqrt{\frac{5(N_c+3)}{N_c^3}}$ 
   \\ 
   $N_{5/2}'$ 
   & $\frac{1}{N_c}$ & $\frac{2}{N_c}$ & $\frac{1}{2N_c}$ &$\frac{1}{6N_c}$   
    \\ 
  $\Delta_{1/2}$
 & $-\frac{4}{3N_c}$ & $\frac{2}{N_c}$ & $-\frac{1}{N_c}$ & 0 
    \\ 
  $\Delta_{3/2}$ 
 & $\frac{2}{3N_c}$ & $\frac{2}{N_c}$ & $-\frac{1}{N_c}$ & 0
   \\ [.2cm]
  \hline 
  \hline
  \end{tabular}
  \caption{Mass operator basis and matrix elements for the  $[\mathbf{20}, 1^-]$   multiplet. }
\label{201minusbasis}}
\end{table}

   \subsection{Operator basis and matrix elements for the  $[{\mathbf{70}}, 1^-]$ multiplet.}

 \begin{table}[b!!]
\renewcommand{\arraystretch}{.75}
 {\squeezetable \footnotesize
    \begin{tabular}{ c c c c c c c }
   \hline
   \hline  & $O_1$ & $O_2$ &~~ $O_3$ &~~ $O_4$ & $O_5$ & $O_6$ \\ 
    & $N_c1$ & $\ell_i\,s_i$ &~~ $\frac{3}{N_c}\ell_{ij}^{(2)}g_{ia}G_{ja}^c$ &~~ $\frac{4}{N_c+1}\,\ell_i\,t_aG_{ja}^c$ & $\frac{1}{N_c}\,\ell_i\,S_i^c$ & $\frac{1}{N_c}S_i^cS_i^c$ \\ [.2cm]
   \hline $^28_{1/2}$ & $N_c$ & $\frac{3-2N_c}{3N_c}$ &~~ 0 &~~ $\frac{2(N_c+3)(3N_c-2)}{9N_c(N_c+1)}$ & $-\frac{N_c+3}{3N_c^2}$ & $\frac{N_c+3}{2N_c^2}$ \\ 
   $^48_{1/2}$ & $N_c$ & $-\frac{5}{6}$ &~~ $-\frac{5(3N_c+1)}{48N_c}$ &~~ $\frac{5(3N_c+1)}{18(N_c+1)}$ & $-\frac{5}{3N_c}$ & $\frac{2}{N_c}$  \\ 
   $^28_{1/2}$ $-$ $^48_{1/2}$ & 0 & $-\sqrt{\frac{(N_c+3)}{18N_c}}$ &~~ $-\frac{5}{24}\sqrt{\frac{(N_c+3)(3N_c-2)^2}{2N_c^3}}$ &~~ $-\frac{(5-3N_c)}{9(N_c+1)}\sqrt{\frac{N_c+3}{2N_c}}$ & $\sqrt{\frac{N_c+3}{18N_c^3}}$ & 0 \\ 
   $^21_{1/2}$ & $N_c$ & -1 &~~ 0 &~~ 0 & 0 & 0 \\ 
    $^210_{1/2}$ & $N_c$ & $\frac{1}{3}$ &~~ 0 &~~ $-\frac{(3N_c+7)}{9(N_c+1)}$ & $-\frac{4}{3N_c}$ & $\frac{2}{N_c}$ \\ 
    $^28_{3/2}$ & $N_c$ & $\frac{2N_c-3}{6N_c}$ &~~ 0 &~~ $-\frac{(N_c+3)(3N_c-2)}{9N_c(N_c+1)}$ & $\frac{N_c+3}{6N_c^2}$ & $\frac{N_c+3}{2N_c^2}$  \\ 
  $^48_{3/2}$ & $N_c$ & $-\frac{1}{3}$ &~~ $\frac{3N_c+1}{12N_c}$ &~~ $\frac{3N_c+1}{9(N_c+1)}$ & $-\frac{2}{3N_c}$ & $\frac{2}{N_c}$ \\ 
    $^28_{3/2}$ $-$ $^48_{3/2}$ & 0 & $-\sqrt{\frac{5(N_c+3)}{36N_c}}$ &~~ $-\frac{1}{48}\sqrt{\frac{5(N_c+3)(2-3N_c)^2}{N_c^3}}$ &~~ $-\sqrt{\frac{5(N_c+3)(5-3N_c)^2}{324N_c(N_c+1)^2}}$ & $\sqrt{\frac{5(N_c+3)}{36N_c^3}}$ & 0 \\ 
   $^21_{3/2}$ & $N_c$ & $\frac{1}{2}$ &~~ 0 &~~ 0 & 0 & 0 \\ 
   $^210_{3/2}$ & $N_c$ & $-\frac{1}{6}$ &~~ 0 &~~ $\frac{3N_c+7}{18(N_c+1)}$ & $\frac{2}{3N_c}$ & $\frac{2}{N_c}$ \\ 
    $^48_{5/2}$ & $N_c$ & $\frac{1}{2}$ &~~ $-\frac{3N_c+1}{48N_c}$ &~~ $-\frac{3N_c+1}{6(N_c+1)}$ & $\frac{1}{N_c}$ & $\frac{2}{N_c}$ \\ [.2cm]
   \hline 
    \hline
   & $O_7$ & $O_8$ & $O_9$ & $O_{10}$ & $O_{11}$ & \\ 
    & $\frac{1}{N_c}s_iS_i^c$ & $\frac{2}{N_c}\ell_{ij}^{(2)}s_iS_j^c$ & $\frac{3}{N_c^2}\,\ell_i\,g_{ja} \{S_j^c,G_{ja}^c\}$ & $\frac{2}{N_c^2}t_a \{S_i^c,G_{ja}^c\}$ & $\frac{3}{N_c^2}\,\ell_i\,g_{ia} \{S_j^c,G_{ja}^c\}$ & \\ [.2cm]
   \hline $^28_{1/2}$ & $-\frac{N_c+3}{4N_c^2}$ & 0 & $\frac{(N_c+3)(7-15N_c)}{24N_c^3}$ & $-\frac{(N_c+3)(3N_c+1)}{12N_c^3}$ & $-\frac{(N_c+3)(3N_c+1)}{24N_c^3}$ & \\ 
   $^48_{1/2}$ & $\frac{1}{2N_c}$ & $\frac{5}{3N_c}$ & $\frac{5(3N_c+1)}{24N_c^2}$ & $-\frac{(3N_c+1)}{3N_c^2}$ & $\frac{5(3N_c+1)}{12N_c^2}$&   \\ 
   $^28_{1/2}$ $-$ $^48_{1/2}$ & 0 & $\sqrt{\frac{25(N_c+3)}{72N_c^3}}$ & $\sqrt{\frac{(N_c+3)(3N_c-2)^2}{288N_c^5}}$ & 0 & $\sqrt{\frac{(N_c+3)(3N_c+1)^2}{72N_c^5}}$& \\ 
   $^21_{1/2}$ & 0 & 0 & 0 & 0 & 0 & \\ 
    $^210_{1/2}$ & $-\frac{1}{N_c}$ & 0 & $\frac{(3N_c+7)}{6N_c^2}$ & $\frac{(3N_c+7)}{6N_c^2}$ & $\frac{(3N_c+7)}{12N_c^2}$ & \\ 
    $^28_{3/2}$ & $-\frac{N_c+3}{4N_c^2}$ & 0 & $\frac{(N_c+3)(15N_c-7)}{48N_c^3}$ & $-\frac{(N_c+3)(3N_c+1)}{12N_c^3}$ & $\frac{(N_c+3)(3N_c+1)}{48N_c^3}$ & \\ 
       $^48_{3/2}$ & $\frac{1}{2N_c}$ & $-\frac{4}{3N_c}$ & $\frac{(3N_c+1)}{12N_c^2}$ & $-\frac{(3N_c+1)}{3N_c^2}$ & $\frac{(3N_c+1)}{6N_c^2}$ &  \\ 
       $^28_{3/2}$ $-$ $^48_{3/2}$ & 0 & $-\sqrt{\frac{5(N_c+3)}{144N_c^3}}$ & $\sqrt{\frac{5(N_c+3)(3N_c-2)^2}{576N_c^5}}$ & 0 & $\sqrt{\frac{5(N_c+3)(3N_c+1)^2}{144N_c^5}}$& \\ 
   $^21_{3/2}$ & 0 & 0 & 0 & 0 & 0 & \\ 
    $^210_{3/2}$ & $-\frac{1}{N_c}$ & 0 & $-\frac{(3N_c+7)}{12N_c^2}$ & $\frac{(3N_c+7)}{6N_c^2}$ & $-\frac{(3N_c+7)}{24N_c^2}$ & \\ 
   $^48_{5/2}$ & $\frac{1}{2N_c}$ & $\frac{1}{3N_c}$ & $-\frac{(3N_c+1)}{8N_c^2}$ & $-\frac{(3N_c+1)}{3N_c^2}$ & $-\frac{(3N_c+1)}{4N_c^2}$ &  \\ [.2cm]
   \hline 
   \end{tabular} \caption{$SU(3)$ singlet basis of operators for the  $[{\mathbf{70}}, 1^-]$ masses. }
   \label{701minusbasis}}
\end{table}
\begin{table}[h!]
\vspace*{-2.1cm}
{\squeezetable
\footnotesize
\renewcommand{\arraystretch}{.45}
\renewcommand{\tabcolsep}{.1cm}
\begin{tabular}{lll}
\hline
\hline \\[-.2cm]
  \multicolumn{1}{c}{}&
\multicolumn{1}{c}{$B_1$}&\multicolumn{1}{c}{$B_2$}\\
&\multicolumn{1}{c}{$t_8$}&\multicolumn{1}{c}{$T_8^c$}\\
 [.2cm]
\hline \\[-.2cm] $^28_{1/2}$,$^28_{3/2}$ & $\frac{N_c^3-(7N_s-8I^2)N_c^2+3(4N_s-8I^2+1)N_c-9N_s}{2\sqrt{3}N_c(N_c-1)(N_c+3)}$ 
&$\frac{N_c^4-(3N_s-1)N_c^3+(N_s-8I^2-3)N_c^2-3(N_s-8I^2+1)N_c+9N_s}{2\sqrt{3}N_c(N_c-1)(N_c+3)}$
 \\ 
  $^48_{1/2}$,$^48_{3/2}$,$^48_{5/2}$ & ~~~$\frac{N_c-N_s-4I^2}{2\sqrt{3}(N_c-1)}$ & $\frac{N_c^2-(3N_s+2)N_c+4(I^2+N_s)}{2\sqrt{3}(N_c-1)}$
   \\ 
    $^28_{1/2}-$$^48_{1/2}$,$^28_{3/2}-$$^48_{3/2}$ & ~~~0 & 0 
    \\ 
     $^21_{1/2}$,$^21_{3/2}$ & ~~~$\frac{(3-N_c)}{\sqrt{3}(N_c+3)}$ & $\frac{(N_c+5)(N_c-3)}{2\sqrt{3}(N_c+3)}$ 
     \\ 
 $^28_{1/2}-$$^21_{1/2}$,$^28_{3/2}-$$^21_{3/2}$ & ~~~$-\frac{3(N_c-1)}{2\sqrt{N_c}(N_c+3)}$ & $-\frac{3(N_c-1)}{2\sqrt{N_c}(N_c+3)}$ 
 \\ 
   $^48_{1/2}-$$^21_{1/2}$,$^48_{3/2}-$$^21_{3/2}$ & ~~~0 & 0 
    \\ 
    $10_{1/2},10_{3/2}$ & ~~~$\frac{N_c-8N_s+5}{2\sqrt{3}(N_c+5)}$ & $\frac{N_c^2-(3N_s-4)N_c-7N_s-5}{2\sqrt{3}(N_c+5)}$ 
     \\ 
 $^28_{1/2}-$$^210_{1/2}$,$^28_{3/2}-$$^210_{3/2}$ & $~~~- \sqrt{\frac{2}{3}\frac{N_c+3}{N_c(N_c-1)(N_c+5)}}$ & $\sqrt{\frac{2}{3}\frac{N_c+3}{N_c(N_c-1)(N_c+5)}}$ 
 \\ 
 $^48_{1/2}-$$^210_{1/2}$,$^48_{3/2}-$$^210_{3/2}$ & ~~~0 & 0 
  \\ [.1cm]
\hline 
\hline
\end{tabular} 
}
{\squeezetable
\footnotesize
\renewcommand{\arraystretch}{.45}
\renewcommand{\tabcolsep}{1.5cm}
\begin{tabular}{l l }
\\[-.2cm] 
  \multicolumn{1}{c}{}&
\multicolumn{1}{c}{$B_3$}\\
  \multicolumn{1}{c}{}&
\multicolumn{1}{c}{ $\frac{10}{N_c}\,d_{8ab}\,g_{ia}G_{ib}^c$ }\\  [.2cm]
\hline \\
 $^28_{1/2}$,$^28_{3/2}$ & $\frac{3N_c^3-(13N_s-8I^2+3)N_c^2+(31N_s-44I^2-12)N_c-6(N_s-14I^2)}{-\frac{24}{5}\sqrt{3}N_c^2(N_c-1)}$
  \\ [-.0003cm]
  $^48_{1/2}$,$^48_{3/2}$,$^48_{5/2}$  & $\frac{3N_c^2-(7N_s+4I^2-3)N_c+(N_s-20I^2)}{-\frac{24}{5}\sqrt{3}N_c(N_c-1)}$\\ 
    $^28_{1/2}-$$^48_{1/2}$,$^28_{3/2}-$$^48_{3/2}$  & 0 
    \\ 
     $^21_{1/2}$,$^21_{3/2}$  & 0 
     \\ 
 $^28_{1/2}-$$^21_{1/2}$,$^28_{3/2}-$$^21_{3/2}$  & $\frac{5(3N_c+1)}{16N_c\sqrt{N_c}}$ 
 \\ 
   $^48_{1/2}-$$^21_{1/2}$,$^48_{3/2}-$$^21_{3/2}$  & 0
    \\ 
    $10_{1/2}$,$10_{3/2}$  & $-\frac{3N_c^2-14(N_s-1)N_c-22N_s-5}{\frac{24}{5}\sqrt{3}N_c(N_c+5)}$
     \\ 
 $^28_{1/2}-$$^210_{1/2}$,$^28_{3/2}-$$^210_{3/2}$ & $\frac{5(N_c+2)}{6\sqrt{6}N_c}\sqrt{\frac{N_c+3}{N_c(N_c-1)(N_c+5)}}$ 
 \\ 
 $^48_{1/2}-$$^210_{1/2}$,$^48_{3/2}-$$^210_{3/2}$  & 0
  \\[.2cm] \hline\hline
\end{tabular} 
}
{\squeezetable
\footnotesize
\renewcommand{\arraystretch}{.45}
\renewcommand{\tabcolsep}{2.2cm}
\begin{tabular}{l l }
  \\[-.3cm] \multicolumn{1}{c}{}&
\multicolumn{1}{c}{$B_4$}\\
  \multicolumn{1}{c}{}&
\multicolumn{1}{c}{ $3\,\ell_i\,g_{i8}$ }\\ [.2cm]
\hline  \\[-.2cm] $^28_{1/2}$ & $-\frac{N_c^3-(10N_s-14I^2+3)N_c^2+3(7N_s-8I^2)N_c-9(N_s-2I^2)}{\sqrt{3}N_c(N_c-1)(N_c+3)}$ \\ 
  $^48_{1/2}$ & $-\frac{5(N_c-N_s-4I^2)}{4\sqrt{3}(N_c-1)}$ \\ 
  $^28_{1/2}-$$^48_{1/2}$ & $-\frac{N_c-N_s-4I^2}{2\sqrt{6}(N_c-1)}\sqrt{1+\frac{3}{N_c}}$ \\ 
   $^21_{1/2}$ & $\frac{\sqrt{3}(N_c-3)}{(N_c+3)}$ \\ 
   $^28_{1/2}-$$^21_{1/2}$ & $\frac{9(N_c-1)}{2(N_c+3)\sqrt{N_c}}$ \\ 
    $^48_{1/2}-^21_{1/2}$ & 0 \\ 
 $^210_{1/2}$ & $\frac{N_c-8N_s+5}{2\sqrt{3}(N_c+5)}$ \\ 
 $^28_{1/2}-$$^210_{1/2}$ & $-\sqrt{\frac{2}{3}}\sqrt{\frac{N_c+3}{N_c(N_c-1)(N_c+5)}}$ \\ 
 $^48_{1/2}-$$^210_{1/2}$ & $\frac{4}{\sqrt{3}}\frac{1}{\sqrt{(N_c-1)(N_c+5)}}$ \\ 
   $^28_{3/2}$ & $\frac{N_c^3-(10N_s-14I^2+3)N_c^2+3(7N_s-8I^2)N_c-9(N_s-2I^2)}{2\sqrt{3}N_c(N_c-1)(N_c+3)}$ \\ 
  $^48_{3/2}$ & $-\frac{N_c-N_s-4I^2}{2\sqrt{3}(N_c-1)}$ \\ 
  $^28_{3/2}-$$^48_{3/2}$ & $-\sqrt{\frac{5}{3}}\frac{N_c-N_s-4I^2}{4(N_c-1)}\sqrt{1+\frac{3}{N_c}}$ \\ 
   $^21_{3/2}$ & $-\frac{\sqrt{3}(N_c-3)}{2(N_c+3)}$ \\ 
   $^28_{3/2}-$$^21_{3/2}$ & $-\frac{9(N_c-1)}{4(N_c+3)\sqrt{N_c}}$ \\ 
   $^48_{3/2}-$$^21_{3/2}$ & 0 \\ 
 $^210_{3/2}$ & $-\frac{N_c-8N_s+5}{4\sqrt{3}(N_c+5)}$ \\ 
 $^28_{3/2}-$$^210_{3/2}$ & $\sqrt{\frac{N_c+3}{6N_c(N_c-1)(N_c+5)}}$ \\ 
    $^48_{3/2}-$$^210_{3/2}$ & $\frac{2\sqrt{10}}{\sqrt{3(N_c-1)(N_c+5)}}$ \\ 
    $^48_{5/2}$ & $\frac{\sqrt{3}(N_c-N_s-4I^2)}{4(N_c-1)}$ \\ [.2cm]
\hline 
\hline
\end{tabular} 
\caption{$SU(3)$ octet basis of operators for the  $[{\mathbf{70}}, 1^-]$ masses.}
\label{701minus8basis}}
\end{table}

\section{Input Masses}

\begin{table}[H]
 \vspace*{2.2cm}
\begin{tabular}{l c c c c   l c c c c}
\hline
\hline 
& && $M_\pi[\text{MeV}]$&&&&$M_\pi[\text{MeV}]$&&\\

Baryon& PDG & 391 & 524 & 702 ~~&~~ Baryon & PDG & 391 & 524 & 702\\ 
\hline $N_{1/2}$ & 938$\pm$30 & 1202$\pm$15 & 1309$\pm$9 & 1473$\pm$4 ~~&~~ $N_{1/2}$ & 1450$\pm$20 & 2221$\pm$52 & 2300$\pm$30 & 2339$\pm$21 \\
 $\Lambda_{1/2}$ & 1116$\pm$30 & 1279$\pm$20 & 1371$\pm$7 & 1473$\pm$4 ~~&~~  $\Lambda_{1/2}$ & 1630$\pm$70 & 2189$\pm$44 & 2330$\pm$26 & 2339$\pm$21 \\ 
 $\Sigma_{1/2}$ & 1189$\pm$30 & 1309$\pm$13 & 1375$\pm$6 & 1473$\pm$4 ~~&~~ $\Sigma_{1/2}$ & 1660$\pm$30 & 2252$\pm$46 & 2357$\pm$52 & 2339$\pm$21 \\ 
 $\Xi_{1/2}$ & 1315$\pm$30 & 1351$\pm$15 & 1420$\pm$9 & 1473$\pm$4 ~~&~~  $\Xi_{1/2}$ & $\cdots $ & 2278$\pm$22 & 2321$\pm$54 & 2339$\pm$21 \\  
\hline $\Delta_{3/2}$ & 1228$\pm$30 & 1518$\pm$20 & 1582$\pm$9 & 1673$\pm$6 ~~&~~ $\Delta_{3/2}$ & 1625$\pm$75 & 2356$\pm$33 & 2450$\pm$17 & 2454$\pm$55 \\ 
 $\Sigma_{3/2}$ & 1383$\pm$30 & 1582$\pm$15 & 1622$\pm$6 & 1673$\pm$6 ~~&~~  $\Sigma_{3/2}$ & $\cdots $ & 2369$\pm$31 & 2423$\pm$19 & 2454$\pm$55 \\ 
 $\Xi_{3/2}$ & 1532$\pm$30 & 1636$\pm$11 & 1655$\pm$11 & 1673$\pm$6 ~~&~~  $\Xi_{3/2}$ & $\cdots $ & 2453$\pm$26 & 2463$\pm$45 & 2454$\pm$55 \\  
 $\Omega_{3/2}$ & 1672$\pm$30 & 1691$\pm$13 & 1694$\pm$9 & 1673$\pm$6 ~~&~~ $\Omega_{3/2}$ & $\cdots $ & 2501$\pm$33 & 2504$\pm$35 & 2454$\pm$55 \\  [.1cm]
\hline \hline
\end{tabular} 
\caption{Ground state (left),  and $[{\bf{56}},0^+]$ excited Roper (right) baryon  masses in MeV. The inversion in the ordering of the masses of the $\Xi_{1/2}$ and the $\Delta$ masses at and above $M_\pi=391$ MeV is similar to that observed in other LQCD calculations \cite{WalkerLoud:2008bp}.}
\label{MassesGS}
\end{table}

 \begin{table}[h!]
 \centering
 \vspace*{2.2cm}
 \begin{tabular}{l c c c c l c c c c}
 \hline
 \hline &&& $M_\pi[\text{MeV}]$&&&&$M_\pi[\text{MeV}]$&&\\
Baryon& PDG & 391 & 524 & 702 & ~~~Baryon & PDG & 391 & 524 & 702\\ 
 \hline $N_{3/2}$ & 1700$\pm$50 & 2148$\pm$33 & 2178$\pm$61 & 2314$\pm$17 &~~  $\Delta_{3/2}$ & 1935$\pm$35 & 2270$\pm$37 & 2344$\pm$17 & 2387$\pm$19 \\ 
  $\Lambda_{3/2}$ & 1800$\pm$30 & 2225$\pm$28 & 2227$\pm$39 & 2314$\pm$17 &~~  $\Sigma_{3/2}''$ & $\cdots $ & 2318$\pm$26 & 2379$\pm$15 & 2387$\pm$19 \\ 
  $\Sigma_{3/2}$ & $\cdots $ & 2243$\pm$24 & 2238$\pm$26 & 2314$\pm$17 &~~ $\Xi_{3/2}''$ & $\cdots $ & 2374$\pm$13 & 2409$\pm$6 & 2387$\pm$19 \\  
  $\Xi_{3/2}$ & $\cdots $ & 2263$\pm$31 & 2305$\pm$15 & 2314$\pm$17 &~~  $\Omega_{3/2}$ & $\cdots $ & 2420$\pm$28 & 2450$\pm$13 & 2387$\pm$19 \\  
 \hline $N_{5/2}$ & 1683$\pm$8 & 2140$\pm$31 & 2198$\pm$17 & 2271$\pm$13 &~~ $\Delta_{5/2}$ & 1895$\pm$25 & 2333$\pm$35 & 2359$\pm$17 & 2388$\pm$17 \\ 
  $\Lambda_{5/2}$ & 1820$\pm$5 & 2228$\pm$20 & 2249$\pm$15 & 2271$\pm$13 &~~ $\Sigma_{5/2}''$ & $\cdots $ & 2368$\pm$20 & 2392$\pm$19 & 2388$\pm$17 \\  
  $\Sigma_{5/2}$ & 1918$\pm$18 & 2229$\pm$22 & 2253$\pm$17 & 2271$\pm$13 &~~  $\Xi_{5/2}''$ & $\cdots $ & 2430$\pm$24 & 2418$\pm$13 & 2388$\pm$17 \\ 
  $\Xi_{5/2}$ & $\cdots $ & 2296$\pm$22 & 2275$\pm$13 & 2271$\pm$13 &~~ $\Omega_{5/2}$ & $\cdots $ & 2487$\pm$24 & 2470$\pm$13 & 2388$\pm$17 \\  
 \hline $\Delta_{1/2}$ & 1895$\pm$25 & 2284$\pm$107 & 2312$\pm$28 & 2398$\pm$32 &~~ $\Delta_{7/2}$ & 1950$\pm$10 & 2390$\pm$31 & 2384$\pm$19 & 2403$\pm$21 \\ 
  $\Sigma_{1/2}''$ & $\cdots $ & 2270$\pm$26 & 2348$\pm$17 & 2398$\pm$32 &~~  $\Sigma_{7/2}''$ & 2033$\pm$8 & 2428$\pm$22 & 2418$\pm$15 & 2403$\pm$21 \\ 
  $\Xi_{1/2}''$ & $\cdots $ & 2293$\pm$35 & 2391$\pm$13 & 2398$\pm$32 &~~  $\Xi_{7/2}''$ & $\cdots $ & 2494$\pm$22 & 2455$\pm$13 & 2403$\pm$21 \\ 
  $\Omega_{1/2}$ & $\cdots $ & 2378$\pm$42 & 2426$\pm$13 & 2398$\pm$32 &~~ $\Omega_{7/2}$ & $\cdots $ & 2553$\pm$22 & 2477$\pm$13 & 2403$\pm$21 \\ 
 [.1cm]
 \hline \hline
 \end{tabular} 
 \caption{$[{\bf{56}},2^+]$ masses. The experimental values are those for baryons  with a three star or higher rating by the PDG.}
 \label{Masses562plus}
 \end{table}

\begin{table}[H]
\centering
 \vspace*{2.2cm}
\begin{tabular}{l c c c c  l c c c c}
\hline
\hline &&& $M_\pi[\text{MeV}]$&&&&$M_\pi[\text{MeV}]$&&\\

Baryon& PDG & 391 & 524 & 702 &~~~~ Baryon  & PDG & 391 & 524 & 702\\ 
\hline $N_{1/2}$ & 1538$\pm$18 & 1681$\pm$51 & 1797$\pm$32 & 1968$\pm$8  &~~~~ $N_{5/2}$ & 1678$\pm$8 & 2012$\pm$26 & 2033$\pm$20 & 2109$\pm$11 \\ 
 $\Lambda_{1/2}$ & 1670$\pm$10 & 1777$\pm$32 & 1852$\pm$27 & 1968$\pm$8 &~~~~ $\Lambda_{5/2}$ & 1820$\pm$10 & 2057$\pm$19 & 2068$\pm$12 & 2109$\pm$11 \\ 
 $\Sigma_{1/2}$ & $\cdots $ & 1783$\pm$25 & 1852$\pm$27 & 1968$\pm$8 &~~~~ $\Sigma_{5/2}$ & 1775$\pm$5 & 2059$\pm$21 & 2066$\pm$15 & 2109$\pm$11 \\  
 $\Xi_{1/2}$ & $\cdots $ & 1846$\pm$32 & 1899$\pm$32 & 1968$\pm$8 &~~~~  $\Xi_{5/2}$ & $\cdots $ & 2127$\pm$21 & 2105$\pm$15 & 2109$\pm$11 \\  
\hline $N_{3/2}$ & 1523$\pm$8 & 1820$\pm$40 & 1896$\pm$17 & 2000$\pm$8 &~~~~  $\Delta_{1/2}$ & 1645$\pm$30 & 1885$\pm$40 & 1964$\pm$42 & 2023$\pm$60 \\  
 $\Lambda_{3/2}$ & 1690$\pm$5 & 1904$\pm$25 & 1939$\pm$17 & 2000$\pm$8 &~~~~  $\Sigma_{1/2}''$ & $\cdots $ & 1952$\pm$25 & 1998$\pm$37 & 2023$\pm$60 \\  
 $\Sigma_{3/2}$ & 1675$\pm$10 & 1905$\pm$23 & 1940$\pm$20 & 2000$\pm$8 &~~~~ $\Xi_{1/2}''$ & $\cdots $ & 1987$\pm$27 & 2038$\pm$17 & 2023$\pm$60 \\
 $\Xi_{3/2}$ & 1823$\pm$5 & 1974$\pm$25 & 1976$\pm$17 & 2000$\pm$8 &~~~~ $\Omega_{1/2}$ & $\cdots $ & 2011$\pm$41 & 2060$\pm$20 & 2023$\pm$60 \\  
\hline $N_{1/2}'$ & 1660$\pm$20 & 1892$\pm$35 & 1928$\pm$37 & 2045$\pm$11 &~~~~  $\Delta_{3/2}$ & 1720$\pm$50 & 1955$\pm$32 & 2033$\pm$17 & 2098$\pm$11 \\  
 $\Lambda_{1/2}'$ & 1785$\pm$65 & 1849$\pm$36 & 1944$\pm$37 & 2045$\pm$11 &~~~~ $\Sigma_{3/2}''$ & $\cdots $ & 1958$\pm$36 & 2071$\pm$15 & 2098$\pm$11 \\ 
 $\Sigma_{1/2}'$ & 1765$\pm$35 & 1840$\pm$36 & 1941$\pm$37 & 2045$\pm$11 &~~~~ $\Xi_{3/2}''$ & $\cdots $ & 2040$\pm$31 & 2108$\pm$15 & 2098$\pm$11 \\ 
 $\Xi_{1/2}'$ & $\cdots $ & 1876$\pm$27 & 2001$\pm$22 & 2045$\pm$11 &~~~~  $\Omega_{3/2}$ & $\cdots $ & 2101$\pm$30 & 2139$\pm$15 & 2098$\pm$11 \\ 
\hline $N_{3/2}'$ & 1700$\pm$50 & 1895$\pm$29 & 1935$\pm$37 & 2077$\pm$10  &~~~~ $\Lambda_{1/2}''$ & 1407$\pm$4 & 1710$\pm$32 & 1796$\pm$20 & 1922$\pm$11 \\ 
 $\Lambda_{3/2}'$ & $\cdots $ & 1936$\pm$30 & 1981$\pm$27 & 2077$\pm$10 &~~~~  $\Lambda_{3/2}''$ & 1520$\pm$1 & 1817$\pm$21 & 1816$\pm$40 & 1903$\pm$11 \\ 
 $\Sigma_{3/2}'$ & $\cdots $ & 1951$\pm$27 & 1977$\pm$25 & 2077$\pm$10 &&&&&\\ 
 $\Xi_{3/2}'$ & $\cdots $ & 1998$\pm$31 & 2030$\pm$27 & 2077$\pm$10 &&&&&\\ [.1cm]
\hline 
\hline  
\end{tabular} 
\caption{$[{\bf{70}},1^-]$   masses.  The experimental values are those for baryons with a three star or higher rating by the PDG.}
\label{Masses701minus}
\end{table}

\newpage
\bibliography{Refs}

\begin{thebibliography}{40}
\expandafter\ifx\csname natexlab\endcsname\relax\def\natexlab#1{#1}\fi
\expandafter\ifx\csname bibnamefont\endcsname\relax
  \def\bibnamefont#1{#1}\fi
\expandafter\ifx\csname bibfnamefont\endcsname\relax
  \def\bibfnamefont#1{#1}\fi
\expandafter\ifx\csname citenamefont\endcsname\relax
  \def\citenamefont#1{#1}\fi
\expandafter\ifx\csname url\endcsname\relax
  \def\url#1{\texttt{#1}}\fi
\expandafter\ifx\csname urlprefix\endcsname\relax\def\urlprefix{URL }\fi
\providecommand{\bibinfo}[2]{#2}
\providecommand{\eprint}[2][]{\url{#2}}

\bibitem[{\citenamefont{Walker-Loud et~al.}(2009)\citenamefont{Walker-Loud,
  Lin, Richards, Edwards, Engelhardt et~al.}}]{WalkerLoud:2008bp}
\bibinfo{author}{\bibfnamefont{A.}~\bibnamefont{Walker-Loud}},
  \bibinfo{author}{\bibfnamefont{H.-W.} \bibnamefont{Lin}},
  \bibinfo{author}{\bibfnamefont{D.}~\bibnamefont{Richards}},
  \bibinfo{author}{\bibfnamefont{R.}~\bibnamefont{Edwards}},
  \bibinfo{author}{\bibfnamefont{M.}~\bibnamefont{Engelhardt}},
  \bibnamefont{et~al.}, \bibinfo{journal}{Phys.Rev.}
  \textbf{\bibinfo{volume}{D79}}, \bibinfo{pages}{054502}
  (\bibinfo{year}{2009}), \eprint{0806.4549}.

\bibitem[{\citenamefont{Bulava et~al.}(2009)\citenamefont{Bulava, Edwards,
  Engelson, Foley, Joo et~al.}}]{Bulava:2009jb}
\bibinfo{author}{\bibfnamefont{J.~M.} \bibnamefont{Bulava}},
  \bibinfo{author}{\bibfnamefont{R.~G.} \bibnamefont{Edwards}},
  \bibinfo{author}{\bibfnamefont{E.}~\bibnamefont{Engelson}},
  \bibinfo{author}{\bibfnamefont{J.}~\bibnamefont{Foley}},
  \bibinfo{author}{\bibfnamefont{B.}~\bibnamefont{Joo}}, \bibnamefont{et~al.},
  \bibinfo{journal}{Phys.Rev.} \textbf{\bibinfo{volume}{D79}},
  \bibinfo{pages}{034505} (\bibinfo{year}{2009}), \eprint{0901.0027}.

\bibitem[{\citenamefont{Edwards et~al.}(2011)\citenamefont{Edwards, Dudek,
  Richards, and Wallace}}]{Edwards:2011jj}
\bibinfo{author}{\bibfnamefont{R.~G.} \bibnamefont{Edwards}},
  \bibinfo{author}{\bibfnamefont{J.~J.} \bibnamefont{Dudek}},
  \bibinfo{author}{\bibfnamefont{D.~G.} \bibnamefont{Richards}},
  \bibnamefont{and} \bibinfo{author}{\bibfnamefont{S.~J.}
  \bibnamefont{Wallace}}, \bibinfo{journal}{Phys.Rev.}
  \textbf{\bibinfo{volume}{D84}}, \bibinfo{pages}{074508}
  (\bibinfo{year}{2011}), \eprint{1104.5152}.

\bibitem[{\citenamefont{Edwards et~al.}(2013)\citenamefont{Edwards, Mathur,
  Richards, and Wallace}}]{Edwards:2012fx}
\bibinfo{author}{\bibfnamefont{R.~G.} \bibnamefont{Edwards}},
  \bibinfo{author}{\bibfnamefont{N.}~\bibnamefont{Mathur}},
  \bibinfo{author}{\bibfnamefont{D.~G.} \bibnamefont{Richards}},
  \bibnamefont{and} \bibinfo{author}{\bibfnamefont{S.~J.}
  \bibnamefont{Wallace}} (\bibinfo{collaboration}{Hadron Spectrum
  Collaboration}), \bibinfo{journal}{Phys.Rev.} \textbf{\bibinfo{volume}{D87}},
  \bibinfo{pages}{054506} (\bibinfo{year}{2013}), \eprint{1212.5236}.

\bibitem[{\citenamefont{Lin et~al.}(2009)}]{Lin:2008pr}
\bibinfo{author}{\bibfnamefont{H.-W.} \bibnamefont{Lin}} \bibnamefont{et~al.}
  (\bibinfo{collaboration}{Hadron Spectrum Collaboration}),
  \bibinfo{journal}{Phys.Rev.} \textbf{\bibinfo{volume}{D79}},
  \bibinfo{pages}{034502} (\bibinfo{year}{2009}), \eprint{0810.3588}.

\bibitem[{\citenamefont{Dudek et~al.}(2013)\citenamefont{Dudek, Edwards, and
  Thomas}}]{Dudek:2012xn}
\bibinfo{author}{\bibfnamefont{J.~J.} \bibnamefont{Dudek}},
  \bibinfo{author}{\bibfnamefont{R.~G.} \bibnamefont{Edwards}},
  \bibnamefont{and} \bibinfo{author}{\bibfnamefont{C.~E.}
  \bibnamefont{Thomas}}, \bibinfo{journal}{Phys.Rev.}
  \textbf{\bibinfo{volume}{D87}}, \bibinfo{pages}{034505}
  (\bibinfo{year}{2013}), \eprint{1212.0830, and references therein.}

\bibitem[{\citenamefont{Bernard et~al.}(2008)\citenamefont{Bernard, Meissner,
  and Rusetsky}}]{Bernard:2007cm}
\bibinfo{author}{\bibfnamefont{V.}~\bibnamefont{Bernard}},
  \bibinfo{author}{\bibfnamefont{U.-G.} \bibnamefont{Meissner}},
  \bibnamefont{and} \bibinfo{author}{\bibfnamefont{A.}~\bibnamefont{Rusetsky}},
  \bibinfo{journal}{Nucl.Phys.} \textbf{\bibinfo{volume}{B788}},
  \bibinfo{pages}{1} (\bibinfo{year}{2008}), \eprint{hep-lat/0702012}.

\bibitem[{\citenamefont{Engel et~al.}(2013)\citenamefont{Engel, Lang, Mohler,
  and Schäfer}}]{Engel:2013ig}
\bibinfo{author}{\bibfnamefont{G.~P.} \bibnamefont{Engel}},
  \bibinfo{author}{\bibfnamefont{C.}~\bibnamefont{Lang}},
  \bibinfo{author}{\bibfnamefont{D.}~\bibnamefont{Mohler}}, \bibnamefont{and}
  \bibinfo{author}{\bibfnamefont{A.}~\bibnamefont{Schäfer}}
  (\bibinfo{collaboration}{BGR}), \bibinfo{journal}{Phys.Rev.}
  \textbf{\bibinfo{volume}{D87}}, \bibinfo{pages}{074504}
  (\bibinfo{year}{2013}), \eprint{1301.4318}.

\bibitem[{\citenamefont{Alexandrou et~al.}(2014)\citenamefont{Alexandrou,
  Korzec, Koutsou, and Leontiou}}]{Alexandrou:2013fsu}
\bibinfo{author}{\bibfnamefont{C.}~\bibnamefont{Alexandrou}},
  \bibinfo{author}{\bibfnamefont{T.}~\bibnamefont{Korzec}},
  \bibinfo{author}{\bibfnamefont{G.}~\bibnamefont{Koutsou}}, \bibnamefont{and}
  \bibinfo{author}{\bibfnamefont{T.}~\bibnamefont{Leontiou}},
  \bibinfo{journal}{Phys.Rev.} \textbf{\bibinfo{volume}{D89}},
  \bibinfo{pages}{034502} (\bibinfo{year}{2014}), \eprint{1302.4410}.

\bibitem[{\citenamefont{Mahbub et~al.}(2013{\natexlab{a}})\citenamefont{Mahbub,
  Kamleh, Leinweber, Moran, and Williams}}]{Mahbub:2013ala}
\bibinfo{author}{\bibfnamefont{M.~S.} \bibnamefont{Mahbub}},
  \bibinfo{author}{\bibfnamefont{W.}~\bibnamefont{Kamleh}},
  \bibinfo{author}{\bibfnamefont{D.~B.} \bibnamefont{Leinweber}},
  \bibinfo{author}{\bibfnamefont{P.~J.} \bibnamefont{Moran}}, \bibnamefont{and}
  \bibinfo{author}{\bibfnamefont{A.~G.} \bibnamefont{Williams}},
  \bibinfo{journal}{Phys.Rev.} \textbf{\bibinfo{volume}{D87}},
  \bibinfo{pages}{094506} (\bibinfo{year}{2013}{\natexlab{a}}),
  \eprint{1302.2987}.

\bibitem[{\citenamefont{Mahbub et~al.}(2013{\natexlab{b}})\citenamefont{Mahbub,
  Kamleh, Leinweber, Moran, and Williams}}]{Mahbub:2012ri}
\bibinfo{author}{\bibfnamefont{M.~S.} \bibnamefont{Mahbub}},
  \bibinfo{author}{\bibfnamefont{W.}~\bibnamefont{Kamleh}},
  \bibinfo{author}{\bibfnamefont{D.~B.} \bibnamefont{Leinweber}},
  \bibinfo{author}{\bibfnamefont{P.~J.} \bibnamefont{Moran}}, \bibnamefont{and}
  \bibinfo{author}{\bibfnamefont{A.~G.} \bibnamefont{Williams}}
  (\bibinfo{collaboration}{CSSM Lattice Collaboration}),
  \bibinfo{journal}{Phys.Rev.} \textbf{\bibinfo{volume}{D87}},
  \bibinfo{pages}{011501} (\bibinfo{year}{2013}{\natexlab{b}}),
  \eprint{1209.0240}.

\bibitem[{\citenamefont{Matagne and Stancu}(2014)}]{Matagne:2014lla}
\bibinfo{author}{\bibfnamefont{N.}~\bibnamefont{Matagne}} \bibnamefont{and}
  \bibinfo{author}{\bibfnamefont{F.}~\bibnamefont{Stancu}}
  (\bibinfo{year}{2014}), \eprint{1406.1791, and references therein}.

\bibitem[{\citenamefont{Gervais and
  Sakita}(1984{\natexlab{a}})}]{Gervais:1983wq}
\bibinfo{author}{\bibfnamefont{J.-L.} \bibnamefont{Gervais}} \bibnamefont{and}
  \bibinfo{author}{\bibfnamefont{B.}~\bibnamefont{Sakita}},
  \bibinfo{journal}{Phys.Rev.Lett.} \textbf{\bibinfo{volume}{52}},
  \bibinfo{pages}{87} (\bibinfo{year}{1984}{\natexlab{a}}).

\bibitem[{\citenamefont{Gervais and
  Sakita}(1984{\natexlab{b}})}]{Gervais:1984rc}
\bibinfo{author}{\bibfnamefont{J.-L.} \bibnamefont{Gervais}} \bibnamefont{and}
  \bibinfo{author}{\bibfnamefont{B.}~\bibnamefont{Sakita}},
  \bibinfo{journal}{Phys.Rev.} \textbf{\bibinfo{volume}{D30}},
  \bibinfo{pages}{1795} (\bibinfo{year}{1984}{\natexlab{b}}).

\bibitem[{\citenamefont{Dashen and Manohar}(1993)}]{Dashen:1993as}
\bibinfo{author}{\bibfnamefont{R.~F.} \bibnamefont{Dashen}} \bibnamefont{and}
  \bibinfo{author}{\bibfnamefont{A.~V.} \bibnamefont{Manohar}},
  \bibinfo{journal}{Phys.Lett.} \textbf{\bibinfo{volume}{B315}},
  \bibinfo{pages}{425} (\bibinfo{year}{1993}), \eprint{hep-ph/9307241}.

\bibitem[{\citenamefont{Goity}(1997)}]{Goity:1996hk}
\bibinfo{author}{\bibfnamefont{J.~L.} \bibnamefont{Goity}},
  \bibinfo{journal}{Phys.Lett.} \textbf{\bibinfo{volume}{B414}},
  \bibinfo{pages}{140} (\bibinfo{year}{1997}), \eprint{hep-ph/9612252}.

\bibitem[{\citenamefont{Pirjol and Yan}(1998)}]{Pirjol:1997sr}
\bibinfo{author}{\bibfnamefont{D.}~\bibnamefont{Pirjol}} \bibnamefont{and}
  \bibinfo{author}{\bibfnamefont{T.-M.} \bibnamefont{Yan}},
  \bibinfo{journal}{Phys.Rev.} \textbf{\bibinfo{volume}{D57}},
  \bibinfo{pages}{5434} (\bibinfo{year}{1998}), \eprint{hep-ph/9711201}.

\bibitem[{\citenamefont{Carlson et~al.}(1998)\citenamefont{Carlson, Carone,
  Goity, and Lebed}}]{Carlson:1998gw}
\bibinfo{author}{\bibfnamefont{C.~E.} \bibnamefont{Carlson}},
  \bibinfo{author}{\bibfnamefont{C.~D.} \bibnamefont{Carone}},
  \bibinfo{author}{\bibfnamefont{J.~L.} \bibnamefont{Goity}}, \bibnamefont{and}
  \bibinfo{author}{\bibfnamefont{R.~F.} \bibnamefont{Lebed}},
  \bibinfo{journal}{Phys.Lett.} \textbf{\bibinfo{volume}{B438}},
  \bibinfo{pages}{327} (\bibinfo{year}{1998}), \eprint{hep-ph/9807334}.

\bibitem[{\citenamefont{Carlson et~al.}(1999)\citenamefont{Carlson, Carone,
  Goity, and Lebed}}]{Carlson:1998vx}
\bibinfo{author}{\bibfnamefont{C.~E.} \bibnamefont{Carlson}},
  \bibinfo{author}{\bibfnamefont{C.~D.} \bibnamefont{Carone}},
  \bibinfo{author}{\bibfnamefont{J.~L.} \bibnamefont{Goity}}, \bibnamefont{and}
  \bibinfo{author}{\bibfnamefont{R.~F.} \bibnamefont{Lebed}},
  \bibinfo{journal}{Phys.Rev.} \textbf{\bibinfo{volume}{D59}},
  \bibinfo{pages}{114008} (\bibinfo{year}{1999}), \eprint{hep-ph/9812440}.

\bibitem[{\citenamefont{Schat et~al.}(2002)\citenamefont{Schat, Goity, and
  Scoccola}}]{Schat:2001xr}
\bibinfo{author}{\bibfnamefont{C.~L.} \bibnamefont{Schat}},
  \bibinfo{author}{\bibfnamefont{J.~L.} \bibnamefont{Goity}}, \bibnamefont{and}
  \bibinfo{author}{\bibfnamefont{N.~N.} \bibnamefont{Scoccola}},
  \bibinfo{journal}{Phys.Rev.Lett.} \textbf{\bibinfo{volume}{88}},
  \bibinfo{pages}{102002} (\bibinfo{year}{2002}), \eprint{hep-ph/0111082}.

\bibitem[{\citenamefont{Goity et~al.}(2002)\citenamefont{Goity, Schat, and
  Scoccola}}]{Goity:2002pu}
\bibinfo{author}{\bibfnamefont{J.~L.} \bibnamefont{Goity}},
  \bibinfo{author}{\bibfnamefont{C.~L.} \bibnamefont{Schat}}, \bibnamefont{and}
  \bibinfo{author}{\bibfnamefont{N.~N.} \bibnamefont{Scoccola}},
  \bibinfo{journal}{Phys.Rev.} \textbf{\bibinfo{volume}{D66}},
  \bibinfo{pages}{114014} (\bibinfo{year}{2002}), \eprint{hep-ph/0209174}.

\bibitem[{\citenamefont{Goity et~al.}(2003{\natexlab{a}})\citenamefont{Goity,
  Schat, and Scoccola}}]{Goity:2003ab}
\bibinfo{author}{\bibfnamefont{J.~L.} \bibnamefont{Goity}},
  \bibinfo{author}{\bibfnamefont{C.~L.} \bibnamefont{Schat}}, \bibnamefont{and}
  \bibinfo{author}{\bibfnamefont{N.~N.} \bibnamefont{Scoccola}},
  \bibinfo{journal}{Phys.Lett.} \textbf{\bibinfo{volume}{B564}},
  \bibinfo{pages}{83} (\bibinfo{year}{2003}{\natexlab{a}}),
  \eprint{hep-ph/0304167}.

\bibitem[{\citenamefont{Pirjol and Schat}(2003)}]{Pirjol:2003ye}
\bibinfo{author}{\bibfnamefont{D.}~\bibnamefont{Pirjol}} \bibnamefont{and}
  \bibinfo{author}{\bibfnamefont{C.}~\bibnamefont{Schat}},
  \bibinfo{journal}{Phys.Rev.} \textbf{\bibinfo{volume}{D67}},
  \bibinfo{pages}{096009} (\bibinfo{year}{2003}), \eprint{hep-ph/0301187}.

\bibitem[{\citenamefont{Gonzalez~de Urreta
  et~al.}(2014)\citenamefont{Gonzalez~de Urreta, Goity, and
  Scoccola}}]{deUrreta:2013koa}
\bibinfo{author}{\bibfnamefont{E.}~\bibnamefont{Gonzalez~de Urreta}},
  \bibinfo{author}{\bibfnamefont{J.~L.} \bibnamefont{Goity}}, \bibnamefont{and}
  \bibinfo{author}{\bibfnamefont{N.~N.} \bibnamefont{Scoccola}},
  \bibinfo{journal}{Phys.Rev.D} \textbf{\bibinfo{volume}{89}},
  \bibinfo{pages}{034024} (\bibinfo{year}{2014}), \eprint{1311.3356}.

\bibitem[{\citenamefont{Matagne and
  Stancu}(2005{\natexlab{a}})}]{Matagne:2004pm}
\bibinfo{author}{\bibfnamefont{N.}~\bibnamefont{Matagne}} \bibnamefont{and}
  \bibinfo{author}{\bibfnamefont{F.}~\bibnamefont{Stancu}},
  \bibinfo{journal}{Phys.Rev.} \textbf{\bibinfo{volume}{D71}},
  \bibinfo{pages}{014010} (\bibinfo{year}{2005}{\natexlab{a}}),
  \eprint{hep-ph/0409261}.

\bibitem[{\citenamefont{Matagne and
  Stancu}(2005{\natexlab{b}})}]{Matagne:2005gd}
\bibinfo{author}{\bibfnamefont{N.}~\bibnamefont{Matagne}} \bibnamefont{and}
  \bibinfo{author}{\bibfnamefont{F.}~\bibnamefont{Stancu}},
  \bibinfo{journal}{Phys.Lett.} \textbf{\bibinfo{volume}{B631}},
  \bibinfo{pages}{7} (\bibinfo{year}{2005}{\natexlab{b}}),
  \eprint{hep-ph/0505118}.

\bibitem[{\citenamefont{Matagne and Stancu}(2006)}]{Matagne:2006zf}
\bibinfo{author}{\bibfnamefont{N.}~\bibnamefont{Matagne}} \bibnamefont{and}
  \bibinfo{author}{\bibfnamefont{F.}~\bibnamefont{Stancu}},
  \bibinfo{journal}{Phys.Rev.} \textbf{\bibinfo{volume}{D74}},
  \bibinfo{pages}{034014} (\bibinfo{year}{2006}), \eprint{hep-ph/0604122}.

\bibitem[{\citenamefont{Matagne and Stancu}(2012)}]{Matagne:2012tm}
\bibinfo{author}{\bibfnamefont{N.}~\bibnamefont{Matagne}} \bibnamefont{and}
  \bibinfo{author}{\bibfnamefont{F.}~\bibnamefont{Stancu}},
  \bibinfo{journal}{Phys.Rev.} \textbf{\bibinfo{volume}{D85}},
  \bibinfo{pages}{116003} (\bibinfo{year}{2012}), \eprint{1205.5207}.

\bibitem[{\citenamefont{Goity}(2005{\natexlab{a}})}]{Goity:2004pw}
\bibinfo{author}{\bibfnamefont{J.~L.} \bibnamefont{Goity}},
  \bibinfo{journal}{Phys.Atom.Nucl.} \textbf{\bibinfo{volume}{68}},
  \bibinfo{pages}{624} (\bibinfo{year}{2005}{\natexlab{a}}),
  \eprint{hep-ph/0405304}.

\bibitem[{\citenamefont{Goity}(2005{\natexlab{b}})}]{Goity:2005gs}
\bibinfo{author}{\bibfnamefont{J.~L.} \bibnamefont{Goity}},
  \bibinfo{journal}{Proceedings of Large Nc QCD 2004, J. L. Goity et al
  Editors. World Scientific, Singapore, 2005.} pp. \bibinfo{pages}{211--222}
  (\bibinfo{year}{2005}{\natexlab{b}}), \eprint{hep-ph/0504121}.

\bibitem[{\citenamefont{Beringer et~al.}(2012)}]{Beringer:1900zz}
\bibinfo{author}{\bibfnamefont{J.}~\bibnamefont{Beringer}} \bibnamefont{et~al.}
  (\bibinfo{collaboration}{Particle Data Group}), \bibinfo{journal}{Phys.Rev.}
  \textbf{\bibinfo{volume}{D86}}, \bibinfo{pages}{010001}
  (\bibinfo{year}{2012}).

\bibitem[{\citenamefont{'t~Hooft}(1974)}]{'tHooft:1973jz}
\bibinfo{author}{\bibfnamefont{G.}~\bibnamefont{'t~Hooft}},
  \bibinfo{journal}{Nucl.Phys.} \textbf{\bibinfo{volume}{B72}},
  \bibinfo{pages}{461} (\bibinfo{year}{1974}).

\bibitem[{\citenamefont{Dashen et~al.}(1995)\citenamefont{Dashen, Jenkins, and
  Manohar}}]{Dashen:1994qi}
\bibinfo{author}{\bibfnamefont{R.~F.} \bibnamefont{Dashen}},
  \bibinfo{author}{\bibfnamefont{E.~E.} \bibnamefont{Jenkins}},
  \bibnamefont{and} \bibinfo{author}{\bibfnamefont{A.~V.}
  \bibnamefont{Manohar}}, \bibinfo{journal}{Phys.Rev.}
  \textbf{\bibinfo{volume}{D51}}, \bibinfo{pages}{3697} (\bibinfo{year}{1995}),
  \eprint{hep-ph/9411234}.

\bibitem[{\citenamefont{Dashen et~al.}(1994)\citenamefont{Dashen, Jenkins, and
  Manohar}}]{Dashen:1993jt}
\bibinfo{author}{\bibfnamefont{R.~F.} \bibnamefont{Dashen}},
  \bibinfo{author}{\bibfnamefont{E.~E.} \bibnamefont{Jenkins}},
  \bibnamefont{and} \bibinfo{author}{\bibfnamefont{A.~V.}
  \bibnamefont{Manohar}}, \bibinfo{journal}{Phys.Rev.}
  \textbf{\bibinfo{volume}{D49}}, \bibinfo{pages}{4713} (\bibinfo{year}{1994}),
  \eprint{hep-ph/9310379}.

\bibitem[{\citenamefont{Goity et~al.}(2003{\natexlab{b}})\citenamefont{Goity,
  Schat, and Scoccola}}]{Goity200383}
\bibinfo{author}{\bibfnamefont{J.~L.} \bibnamefont{Goity}},
  \bibinfo{author}{\bibfnamefont{C.~L.} \bibnamefont{Schat}}, \bibnamefont{and}
  \bibinfo{author}{\bibfnamefont{N.~N.} \bibnamefont{Scoccola}},
  \bibinfo{journal}{Physics Letters B} \textbf{\bibinfo{volume}{564}},
  \bibinfo{pages}{83 } (\bibinfo{year}{2003}{\natexlab{b}}), ISSN
  \bibinfo{issn}{0370-2693}.

\bibitem[{\citenamefont{Jenkins et~al.}(2010)\citenamefont{Jenkins, Manohar,
  Negele, and Walker-Loud}}]{Jenkins:2009wv}
\bibinfo{author}{\bibfnamefont{E.~E.} \bibnamefont{Jenkins}},
  \bibinfo{author}{\bibfnamefont{A.~V.} \bibnamefont{Manohar}},
  \bibinfo{author}{\bibfnamefont{J.~W.} \bibnamefont{Negele}},
  \bibnamefont{and}
  \bibinfo{author}{\bibfnamefont{A.}~\bibnamefont{Walker-Loud}},
  \bibinfo{journal}{Phys.Rev.} \textbf{\bibinfo{volume}{D81}},
  \bibinfo{pages}{014502} (\bibinfo{year}{2010}), \eprint{0907.0529}.

\bibitem[{\citenamefont{Calle~Cordon et~al.}(2014)\citenamefont{Calle~Cordon,
  DeGrand, and Goity}}]{Cordon:2014sda}
\bibinfo{author}{\bibfnamefont{A.}~\bibnamefont{Calle~Cordon}},
  \bibinfo{author}{\bibfnamefont{T.}~\bibnamefont{DeGrand}}, \bibnamefont{and}
  \bibinfo{author}{\bibfnamefont{J.~L.} \bibnamefont{Goity}},
  \bibinfo{journal}{Phys.Rev.} \textbf{\bibinfo{volume}{D90}},
  \bibinfo{pages}{014505} (\bibinfo{year}{2014}), \eprint{1404.2301}.

\bibitem[{\citenamefont{Carlson and Carone}(2000)}]{Carlson:2000zr}
\bibinfo{author}{\bibfnamefont{C.~E.} \bibnamefont{Carlson}} \bibnamefont{and}
  \bibinfo{author}{\bibfnamefont{C.~D.} \bibnamefont{Carone}},
  \bibinfo{journal}{Phys.Lett.} \textbf{\bibinfo{volume}{B484}},
  \bibinfo{pages}{260} (\bibinfo{year}{2000}), \eprint{hep-ph/0005144}.

\bibitem[{\citenamefont{Isgur and Karl}(1978)}]{Isgur:1978xj}
\bibinfo{author}{\bibfnamefont{N.}~\bibnamefont{Isgur}} \bibnamefont{and}
  \bibinfo{author}{\bibfnamefont{G.}~\bibnamefont{Karl}},
  \bibinfo{journal}{Phys.Rev.} \textbf{\bibinfo{volume}{D18}},
  \bibinfo{pages}{4187} (\bibinfo{year}{1978}).

\bibitem[{\citenamefont{Jayalath et~al.}(2011)\citenamefont{Jayalath, Goity,
  Gonzalez~de Urreta, and Scoccola}}]{Jayalath:2011uc}
\bibinfo{author}{\bibfnamefont{C.}~\bibnamefont{Jayalath}},
  \bibinfo{author}{\bibfnamefont{J.~L.} \bibnamefont{Goity}},
  \bibinfo{author}{\bibfnamefont{E.}~\bibnamefont{Gonzalez~de Urreta}},
  \bibnamefont{and} \bibinfo{author}{\bibfnamefont{N.~N.}
  \bibnamefont{Scoccola}}, \bibinfo{journal}{Phys.Rev.}
  \textbf{\bibinfo{volume}{D84}}, \bibinfo{pages}{074012}
  (\bibinfo{year}{2011}), \eprint{1108.2042}.

\end{thebibliography}

\end{document}